\documentclass[12pt]{article}
\usepackage{chet}
\usepackage{booktabs}
\usepackage{amsmath,amssymb,dsfont,mathtools}
\usepackage[square, comma, sort&compress,numbers]{natbib}
\usepackage{graphicx,url}
\usepackage{color}

\usepackage{epsfig}

\newcommand   \be     {\begin{equation}}
\newcommand   \ee     {\end{equation}} 
\newcommand   \bee    {\begin{equation}\begin{aligned}}
\newcommand   \eee    {\end{aligned}\end{equation}}

\def\d{\mathrm{d}}
\def\Re{\mathrm{Re}}
\def\Im{\mathrm{Im}}

\begin{document}

%%%%%%%%%%% Title page
\title{Analytic Bounds and Emergence of $\textrm{AdS}_2$ Physics from the Conformal Bootstrap}
%%%%%%%%%%%
\author{Dalimil Maz\'{a}\v{c}}
%%%%%%%%%%%
\affiliation{
Perimeter Institute for Theoretical Physics, Waterloo, ON N2L 2Y5, Canada\\ \&\\
Department of Physics and Astronomy, University of Waterloo, ON N2L 3G1, Canada
}
\emails{dalimil.mazac@gmail.com}
%%%%%%%%%%%
\abstract{We study analytically the constraints of the conformal bootstrap on the low-lying spectrum of operators in field theories with global conformal symmetry in one and two spacetime dimensions. We introduce a new class of linear functionals acting on the conformal bootstrap equation. In 1D, we use the new basis to construct extremal functionals leading to the optimal upper bound on the gap above identity in the OPE of two identical primary operators of integer or half-integer scaling dimension. We also prove an upper bound on the twist gap in 2D theories with global conformal symmetry. When the external scaling dimensions are large, our functionals provide a direct point of contact between crossing in a 1D CFT and scattering of massive particles in large $\textrm{AdS}_2$. In particular, CFT crossing can be shown to imply that appropriate OPE coefficients exhibit an exponential suppression characteristic of massive bound states, and that the 2D flat-space S-matrix should be analytic away from the real axis.}

\maketitle

\tableofcontents

\flushbottom

\newpage

%%%%%%%%%%%%%%% Section: Introduction
\section{Introduction}\label{sec:introduction}
Unitarity and associativity of the operator product expansion have proven very powerful in constraining the dynamics of conformal field theories (CFTs) in various dimensions. These principles, jointly referred to as the conformal bootstrap, can be used for example to analytically derive universal behaviour of CFTs at large spin \cite{LargeSpin1,LargeSpin2}, the emergence of local physics in the AdS dual \cite{LocalAdS}, Hofman-Maldacena bounds \cite{HMfromCFT} or causality \cite{HartmanCausality}.

Some of the most exciting consequences of the conformal bootstrap equations are constraints on the low-lying spectrum of operators. Most prominently, there is a strong numerical evidence that the 3D Ising model at criticality is the unique 3D CFT with a $\mathbb{Z}_2$ symmetry and precisely one relevant scalar primary operator of each $\mathbb{Z}_2$ charge \cite{3dIsing1,3dIsing2,3dIsingMixed}. In spite of a substantial progress on the numerical front, little has been learnt about the analytic origin of these constraints. The main aim of this article is to take some steps towards such analytic understanding.

A standard example of an equation arising in the conformal bootstrap expresses the crossing symmetry of the four-point function of identical primary operators $\phi(x)$ and takes the form
\be
\sum\limits_{\mathcal{O}\in\phi\times\phi}(c_{\phi\phi\mathcal{O}})^2F_{\mathcal{O}}(z,\bar{z}) = 0\,,
\label{eq:BE1}
\ee
where the sum runs over primary operators present in the $\phi\times\phi$ OPE, $c_{\phi\phi\mathcal{O}}$ is the corresponding OPE coefficient, and $F_{\mathcal{O}}(z,\bar{z})$ are functions related to conformal blocks and completely fixed by conformal symmetry in terms of the quantum numbers of $\phi$ and $\mathcal{O}$ and the dimension of spacetime. Unitarity implies $(c_{\phi\phi\mathcal{O}})^2>0$.

\eqref{eq:BE1} can be looked upon as a vector equation in the infinite-dimensional vector space of functions of two complex variables $z$ and $\bar{z}$. It is mostly due to the infinite-dimensional nature of the problem that an extraction of physical consequences from \eqref{eq:BE1} is not a simple task. The challenge is to identify a direction in this vector space along which the bootstrap equation is the most revealing. Speaking more formally, any linear functional acting on the space of functions $F_{\mathcal{O}}(z,\bar{z})$ can be applied to \eqref{eq:BE1}, leading to a single constraint on the CFT data. Some functionals lead to stronger constraints than others. The functionals leading to optimal constraints have been called extremal functionals \cite{ExtremalFunctional}. The extremal functional depends on the precise question we are asking but can be expected to carry valuable physical information about conformal field theories. An analytic construction of various extremal functionals is therefore a promising strategy for understanding the bootstrap bounds.

One example of a constraint that \eqref{eq:BE1} implies for the CFT data is an upper bound on the gap in the spectrum of scalar $\mathcal{O}$ above identity. This bound exhibits a kink at the critical Ising model both in two and three dimensions, and the two are continuously connected across dimensions \cite{FractionalD}. An analytic derivation of the shape of this bound already in 2D with global conformal symmetry is therefore a very important problem.

In the present paper, we take a step in this direction by finding the optimal upper bound on the gap in one-dimensional theories with global conformal symmetry. Such theories are interesting in their own right since they describe conformal line defects in higher-dimensional CFTs \cite{DefectNumerics,DefectBootstrap}, models of (super)conformal quantum mechanics, as well as field theories placed in $\textrm{AdS}_2$ \cite{SMatrix1}. The conformal bootstrap equations in 1D are relatively simple since the conformal blocks are hypergeometric functions of a single cross-ratio $z$. Moreover, the global conformal blocks in 2D are products of two copies of 1D conformal blocks so one can hope to lift bootstrap results from 1D to 2D.

Article \cite{DefectBootstrap} presented numerical evidence that in unitary 1D CFTs, the optimal upper bound on the scaling dimension of the lowest primary operator above identity in the OPE of two identical primary operators $\psi(x)$ is
\be
\tilde{\Delta} = 2\Delta_\psi + 1
\ee
for any $\Delta_\psi>0$. In fact, the bound can not be any lower since this value is saturated by the boundary correlators of a free massive Majorana fermion in $\textrm{AdS}_2$. Indeed, the primary operators in the $\psi\times\psi$ OPE are the two-particle states $\psi\!\overleftrightarrow{\partial}^{2j+1}\!\psi$, $j\geq 0$, the lowest scaling dimension being $2\Delta_\psi + 1$.

We will prove that $2\Delta_\psi + 1$ is the optimal bound for $\Delta_\psi$ positive integer or half-integer by analytically constructing the corresponding extremal functionals. Traditional numerical bootstrap relies on functionals in the form of linear combinations of derivatives in $z$ evaluated at the crossing-symmetric point $z=1/2$. We will demonstrate that the correct extremal functionals do not lie in the space spanned by this set. Instead, we will introduce a new class of functionals taking the form of integrals of the discontinuity of the conformal blocks on the branch cut $z\in(1,\infty)$ against a suitable integral kernel. The integral kernel corresponding to the extremal functional can be fixed analytically. We checked that the derivative functionals coming from the numerics converge to our analytic functional when expressed in the new basis as we approach the optimal bound.

Thanks to its distinguished nature, the analytic extremal functional $\omega_{\Delta_\psi}$ can be expected to imply important consequences for any 1D CFT. Acting with $\omega_{\Delta_\psi}$ on the equation \eqref{eq:BE1}, we obtain
\be
\sum\limits_{\mathcal{O}\in\psi\times\psi}(c_{\psi\psi\mathcal{O}})^2
\omega_{\Delta_\psi}(F_{\mathcal{O}}(z)) = 0\,.
\label{eq:BE2}
\ee
The free fermion theory trivially satisfies this equation since $\omega_{\Delta_\psi}$ vanishes on the spectrum of the extremal solution. However, \eqref{eq:BE2} represents a universal constraint satisfied by any consistent four-point function. This constraint is particularly revealing for $\Delta_\psi\gg 1$. We will show that a family of unitary solutions of \eqref{eq:BE2} where the dimensions of all primary operators scale linearly with $\Delta_\psi$ as $\Delta_\psi\rightarrow\infty$ has many features of a boundary four-point function corresponding to scattering in a massive QFT placed in large $\textrm{AdS}_2$. Specifically, we will recover the precise exponential supression of OPE coefficients of operators corresponding to bound states seen in \cite{Raul1,Raul2,SMatrix1} and universal behaviour of OPE coefficients corresponding to two-particle states derived in \cite{SMatrix1}. The validity of equation \eqref{eq:BE2} will then be seen to require analyticity of the flat-space S-matrix in the upper-half plane, together with a sum rule for the OPE coefficients of two-particle states at rest.

Finally, we can use the relationship between 1D conformal blocks and 2D global conformal blocks to lift the 1D extremal functionals to closely related functionals acting on the 2D crossing equation. These functionals then imply that the OPE of two identical scalar primaries $\phi(x)$ must contain a non-identity global conformal primary with twist $\tau$ satisfying
\be
\tau \leq 2\Delta_\phi + 2\,.
\ee
This bound is valid without assuming Virasoro symmetry, so also for 2D conformal boundaries and surface defects. Theories with Virasoro symmetry automatically satisfy it thanks to the existence of zero-twist operators other than identity. However, when $0<\Delta_\phi<1$, we can show that the bound must be satisfied by a primary with strictly positive twist, thereby getting a nontrivial prediction also in the presence of Virasoro symmetry.

The rest of this article is organized as follows. Section \ref{sec:review} is a review of ideas useful in the remaining parts, namely extremal functionals and the conformal bootstrap in 1D. We use section \ref{sec:commentsfunctional} to motivate and introduce a new class of 1D bootstrap functionals. In section \ref{sec:extremalfunctionals}, we explain the virtues of the new basis and analytically construct the extremal functional for $\Delta_\psi=1/2$. We extend the construction to other integer and half-integer values of $\Delta_\psi$ in section \ref{sec:higherdelta}. We explain how applying the new functionals at large $\Delta_\psi$ naturally leads to the physics of massive (1+1)D QFTs in large $\textrm{AdS}_2$ in section \ref{sec:ads} and prove an upper bound on the minimal twist in 2D in section \ref{sec:2d}. Future directions are outlined in section \ref{sec:future}.
\newpage
%%%%%%%%% Section: Review
\section{Review}\label{sec:review}
%%%%%%%%%%% Subsection: The conformal bootstrap and extremal functionals
\subsection{The conformal bootstrap and extremal functionals}\label{ssec:functionals}
We start by explaining the basic idea of the conformal bootstrap. See \cite{NotesSlava,NotesDSD} for more complete reviews. The simplest example of constraints that the conformal bootstrap imposes on the low-lying spectrum of primary operators comes from considering the four-point function of a neutral scalar primary operator $\phi(x)$. Thanks to the conformal symmetry, the four-point function takes the form
\be
\langle \phi(x_1)\phi(x_2)\phi(x_3)\phi(x_4)\rangle = \frac{1}{|x_{12}|^{2\Delta_{\phi}}|x_{34}|^{2\Delta_{\phi}}}A(z,\bar{z})\,,
\ee
with $A(z,\bar{z})$ unconstrained by conformal symmetry alone, and where $z$ and $\bar{z}$ are defined by their relation to the conformal cross-ratios
\be
z\bar{z} = \frac{x_{12}^2x_{34}^3}{x_{13}^2x_{24}^2}\,,\quad
(1-z)(1-\bar{z}) = \frac{x_{14}^2x_{23}^3}{x_{13}^2x_{24}^2}\,,
\ee
with $x_{ij}=x_i - x_j$. Applying the operator product expansion (OPE) to $\phi(x_1)\phi(x_2)$ leads to the following expansion of $A(z,\bar{z})$
\be
A(z,\bar{z}) = \sum_{\mathcal{O}\in \phi\times\phi}(c_{\phi \phi \mathcal{O}})^2G_{\Delta_{\mathcal{O}},s_{\mathcal{O}}}(z,\bar{z})\,,
\label{eq:ope1}
\ee
where the sum ranges through primary operators appearing in the $\phi\times\phi$ OPE, which are characterized by their scaling dimension $\Delta_{\mathcal{O}}$ and spin $s_{\mathcal{O}}$. The conformal blocks $G_{\Delta,s}(z,\bar{z})$ are fixed by conformal symmetry in terms of $\Delta$, $s$ and the dimension of spacetime $d$. In unitary theories, $(c_{\phi\phi\mathcal{O}})^2$ has the following interpretation in terms of the scalar product $\langle\cdot|\cdot\rangle$ in the Hilbert space of the theory on $S^{d-1}\times\mathbb{R}$
\be
(c_{\phi \phi \mathcal{O}})^2 = \frac{\langle\phi|\phi(0)|\mathcal{O}\rangle\langle \mathcal{O}|\phi(0)|\phi\rangle}{\langle\mathcal{O}|\mathcal{O}\rangle} = \frac{|\langle \mathcal{O}|\phi(0)|\phi\rangle|^2}{\langle\mathcal{O}|\mathcal{O}\rangle} 
\ee
and thus is positive. We can assume the identity operator appears in the $\phi\times\phi$ OPE. Crucially, we can also apply the OPE to $\phi(x_1)\phi(x_4)$, leading to the expansion
\be
A(z,\bar{z}) =\left[ \frac{z\bar{z}}{(1-z)(1-\bar{z})}\right ]^{\Delta_\phi}
\sum_{\mathcal{O}\in \phi\times\phi}(c_{\phi \phi \mathcal{O}})^2
G_{\Delta_{\mathcal{O}},s_{\mathcal{O}}}(1-z,1-\bar{z})\,.
\label{eq:ope2}
\ee
The consistency of the expansions \eqref{eq:ope1}, \eqref{eq:ope2} can be written more succintly as
\be
\sum_{\mathcal{O}\in \phi\times\phi}(c_{\phi \phi \mathcal{O}})^2 F_{\Delta_{\mathcal{O}},s_{\mathcal{O}}}(z,\bar{z}) = 0\,,
\label{eq:bootstrap}
\ee
where
\be
F_{\Delta,s}(z,\bar{z}) =
(z\bar{z})^{-\Delta_\phi}G_{\Delta,s}(z,\bar{z}) - (z\leftrightarrow 1-z,\bar{z}\leftrightarrow 1-\bar{z})\,.
\ee
Equation \eqref{eq:bootstrap} imposes constraints on the spectrum of primary operators in the $\phi\times\phi$ OPE. Only for certain choices of the spectrum will there exist positive coefficients $(c_{\phi \phi \mathcal{O}})^2$ satisfying \eqref{eq:bootstrap}. $F_{\Delta,s}(z,\bar{z})$ should be thought of as a holomorphic function of two independent complex variables $z$, $\bar{z}$. In each of the variables, it has branch points at $z,\bar{z}=0,1,\infty$, where the branch cuts can be chosen to run from $-\infty$ to $0$ and from $1$ to $\infty$. Equation \eqref{eq:bootstrap} holds everywhere away from these branch cuts. Either of the two OPE expansions stops converging on some of the branch cuts, and consequently it is not legal to analytically continue the equation through the branch cuts. However, the equation holds arbitrarily close to the branch cuts, provided we stay on the first sheet.

The mechanism through which equation \eqref{eq:bootstrap} constrains the spectrum in the $\phi\times\phi$ OPE can be usefully cast in the language of linear functionals $\omega$ acting on the functions $F_{\Delta,s}(z,\bar{z})$. Indeed, suppose we have such functional
\be
\omega: F_{\Delta,s} \mapsto \omega(\Delta,s)\in \mathbb{R}
\ee
and suppose that $\omega$ is non-negative on a candidate spectrum $\mathcal{S}$ of primary operators appearing in the $\phi\times\phi$ OPE
\be
\forall\,(\Delta,s)\in\mathcal{S}:\quad\omega(\Delta,s)\geq 0\,.
\ee
Applying $\omega$ to \eqref{eq:bootstrap} we find that $\mathcal{S}$ can be a consistent spectrum only if $\omega$ vanishes on all of $\mathcal{S}$. Moreover, the converse also holds, namely whenever we have a spectrum $\mathcal{S}$ for which no solution of \eqref{eq:bootstrap} can be found, there is always a functional non-negative on all of $\mathcal{S}$ and strictly positive on at least one operator $(\Delta,s)\in\mathcal{S}$.
%%%%%%%%% Figure illustrating the action of extremal functionals
\begin{figure}
\centering
\includegraphics[width=0.9\textwidth]{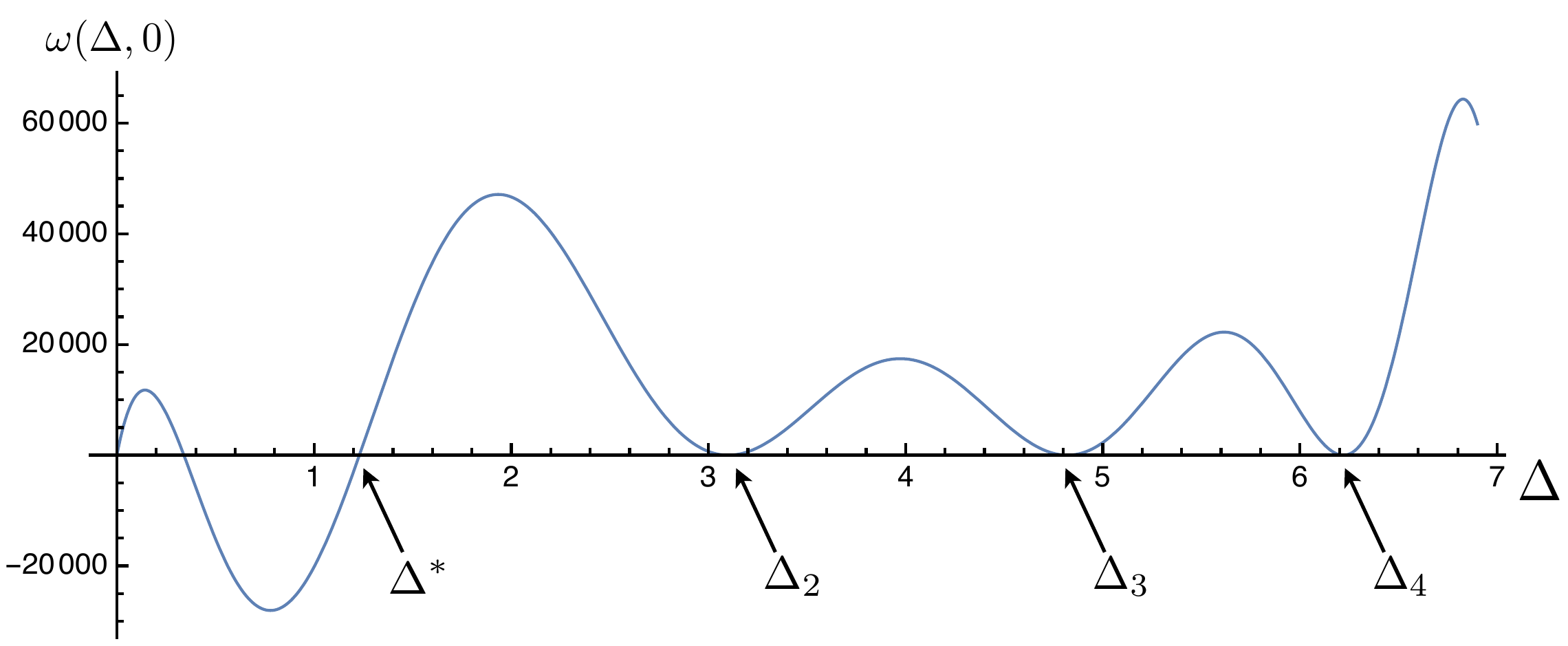}
\caption{The action of a typical extremal functional for the bound on the scalar gap on $F_{\Delta,0}$. The leading non-identity operator appears at a first-order zero with a positive slope, while higher operators lie at second-order zeros.}\label{fig:ExtFun1}
\end{figure}
%%%%%%%%%

In order to search for an upper bound on the gap in the scalar sector above identity without reference to the rest of the spectrum, we should focus on nonzero functionals such that\footnote{In numerical implementations, the first condition is usually replaced by $\omega(0,0) = 1$ in order to avoid the identically zero functional. Such functionals can only become extremal asymptotically, when $\omega(\Delta,s)/\omega(0,0)\rightarrow\infty$ for a generic $(\Delta,s)$ outside of the spectrum.}
\bee
\omega(0,0)&\geq 0\\
\omega(\Delta,0)&\geq 0\quad\textrm{for}\quad\Delta\geq\Delta^*\\
\omega(\Delta,s)&\geq 0\quad\textrm{for}\quad\Delta\geq d+s-2\;\;\textrm{and}\;s\geq 2\,.
\eee
The minimal $\Delta^*$ for which such $\omega$ exists coincides with the upper bound on the scalar gap. We denote this upper bound as $\tilde{\Delta}$. Consider a unitary solution to crossing with gap $\tilde{\Delta}$. As pointed out in \cite{ExtremalFunctional}, all operators in the solution must correspond to zeros of any functional $\omega$ for which $\Delta^* = \tilde{\Delta}$. Functionals for which $\Delta^*=\tilde{\Delta}$ are called extremal functionals. Figure \ref{fig:ExtFun1} illustrates how a typical extremal functional corresponding to the upper bound on the scalar gap acts on $F_{\Delta,0}$. It vanishes at $\Delta=0$ and has a first-order zero and positive slope at the lowest non-identity operator with dimension $\Delta^*$. The functional must be negative immediately to the left of $\Delta^*$ since otherwise it would not exclude solutions with gap smaller than $\Delta^*$. Higher-lying scalar operators in the spectrum sit at second-order zeros since the functional must vanish there without ever becoming negative for $\Delta>\Delta^*$.

Generically, we expect both the extremal functional and the corresponding extremal solution of \eqref{eq:bootstrap} to be unique up to an overall positive rescaling. One counterexample is the free theory point in 4D, i.e. $\Delta_\phi=1$, where an infinite class of extremal functionals leads to the unique free theory solution. The extremal functionals for the 1D bound studied in this paper will be unique up to an overall rescaling. There is no reason for all the zeros of the extremal functional to correspond to operators appearing in the solution to crossing with nonzero OPE coefficient. A typical example is the first first-order zero in Figure \ref{fig:ExtFun1} with negative slope, but sometimes even spurious second-order zeros can occur above $\Delta^*$. We will find that the extremal functionals for 1D bootstrap do not contain such subtleties.

One should view the extremal functional as the optimal lens with which to study the bootstrap equation. It is the functional that projects the infinite-dimensional bootstrap equation on a one-dimensional space in the most revealing manner. It is likely that understanding the mechanism through which the conformal bootstrap leads to bounds on the gap, features in these bounds as well as islands in multi-correlator bootstrap amounts to understanding the precise nature of the extremal functionals. In this paper, we will also see that extremal functionals carry valuable physical information about solutions to crossing distinct from the extremal solution. Indeed, the extremal functionals for the 1D bootstrap bound will be shown to naturally lead to the physics of QFT in $\textrm{AdS}_2$ of large radius when the external scaling dimensions are large.

%%%%%%%%%%% Subsection: The conformal bootstrap in one dimension
\subsection{The conformal bootstrap in one dimension}\label{ssec:1Dbootstrap}
There are good reasons to start an analytic study of the constraining power of the bootstrap equations in one spacetime dimension. The kinematics is very simple, and explicit formulas exist for arbitrary conformal blocks. Moreover, one can hope to lift the 1D results to 2D, where the conformal blocks are linear combinations of products of the 1D blocks. Finally, as we will review shortly, an explicit formula likely exists for the optimal 1D bootstrap bound, begging for an analytic explanation. Numerous interesting systems exhibit the global conformal symmetry in one dimension, including conformal boundaries in 2D CFTs, line defects in general CFTs \cite{DefectNumerics,DefectBootstrap}, and various examples of $\mathrm{AdS_2/CFT_1}$ holography.

Here and in the rest of the article, by conformal symmetry we always mean the global conformal symmetry. In one dimension, the conformal group is $SL(2)$, with generators $D,P,K$ satisfying commutation relations
\be
[D,P] = P\,,\quad [D,K] = - K\,,\quad [K,P] = 2 D\,.
\ee
Unitary highest-weight representations, corresponding to primary fields, are labelled by the scaling dimension $\Delta$. There are no rotations and therefore no spin. Two- and three-point functions are completely fixed in terms of $\Delta_i$ and structure constants $c_{ijk}.$\footnote{Note that unlike in higer dimensions, $c_{ijk}\neq c_{jik}$ in general because two operators can not be continuously swapped in 1D. However, we still expect $c_{ijk}=c_{jki}$ since a line is conformally equivalent to a circle.}
Four points on a line give rise to a single cross-ratio
\be
z = \frac{x_{12}x_{34}}{x_{13}x_{24}}\,.
\ee
We can focus on the kinematic region where $x_1<x_2<x_3<x_4$ and use the three conformal generators to set $x_1=0$, $x_3=1$, $x_4=\infty$, so that $x_2=z\in(0,1)$. The four-point function of identical primary fields $\psi(x)$ takes the form
\be
\langle \psi(x_1)\psi(x_2)\psi(x_3)\psi(x_4)\rangle = \frac{1}{|x_{12}|^{2\Delta_{\psi}}|x_{34}|^{2\Delta_{\psi}}}A(z)\,,
\ee
where $A(z)$ can be expanded in conformal blocks
\be
A(z) = \sum\limits_{\mathcal{O}\in\psi\times\psi}(c_{\psi\psi\mathcal{O}})^2G_{\Delta_{\mathcal{O}}}(z)\,.
\ee
The 1D conformal block is just the chiral half of the 2D global conformal block \cite{DolanOsborn1}
\be
G_{\Delta}(z) = z^{\Delta}{}_2F_1(\Delta,\Delta;2\Delta;z)\,.
\label{eq:block1D}
\ee
Assuming $\psi(x)$ is a real field, the conformal block expansion starts with the identity operator with $\Delta=0$. The crossing equation reads
\be
\sum\limits_{\mathcal{O}\in\psi\times\psi}(c_{\psi\psi\mathcal{O}})^2F_{\Delta_{\mathcal{O}}}(z)=0\,,
\label{eq:be1D}
\ee
where
\be
F_{\Delta}(z) = z^{-2\Delta_\psi}G_{\Delta}(z) -  (1-z)^{-2\Delta_\psi}G_{\Delta}(1-z)\,.
\label{eq:functionbasis}
\ee

\begin{figure}
\centering
\includegraphics[width=0.7\textwidth]{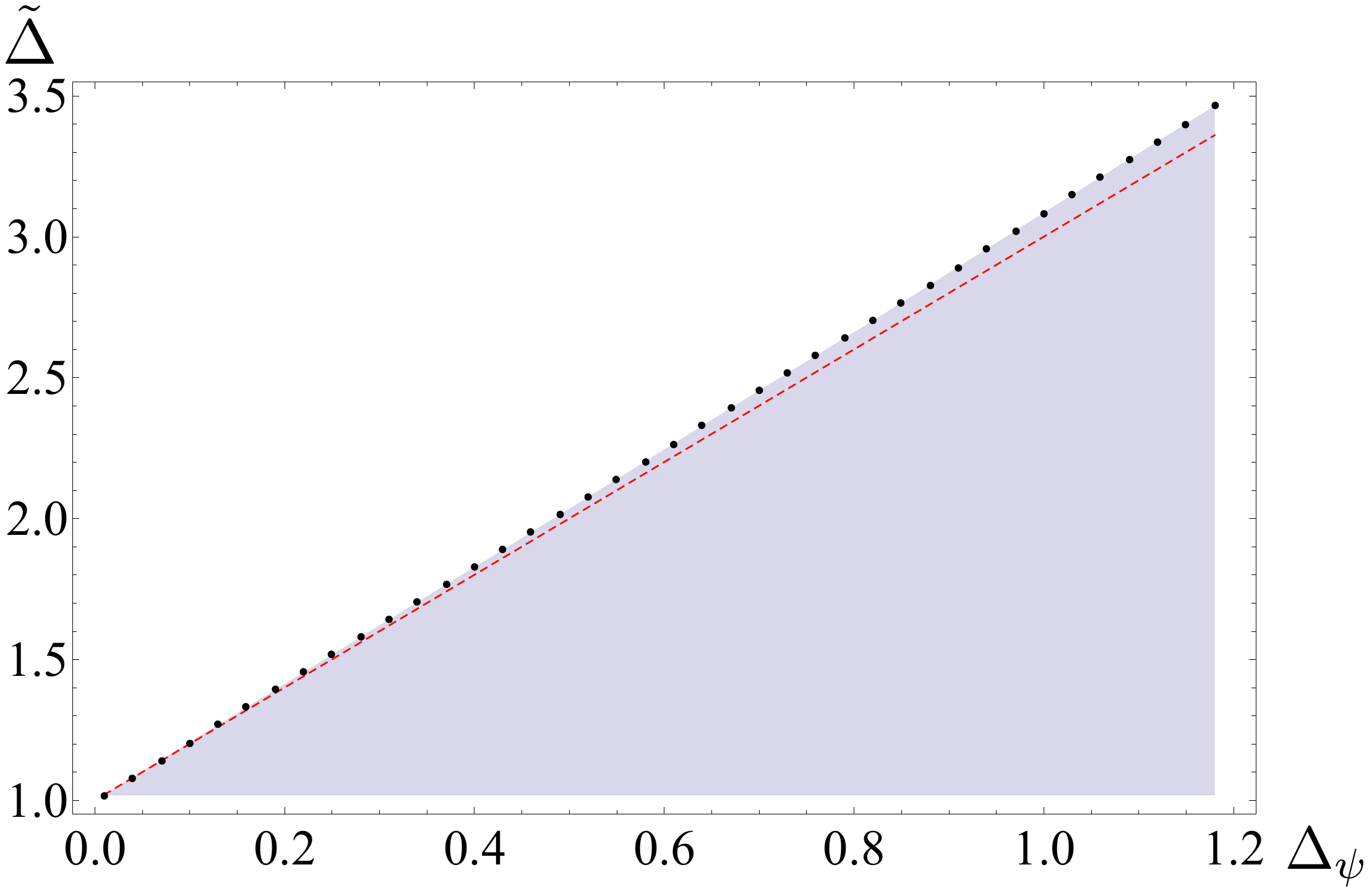}
\caption{Black dots: Numerical bootstrap bound on the gap above identity following from \eqref{eq:be1D}, using 50 derivatives. Red dashed line: $\tilde{\Delta}=2\Delta_\psi + 1$. Figure taken from \cite{DefectBootstrap}.}\label{fig:1DBound}
\end{figure}

Standard numerical bootstrap applied to \eqref{eq:be1D} using derivatives at the crossing-symmetric point $z=1/2$ leads to an upper bound on the scaling dimension of the first non-identity operator in the $\psi\times\psi$ OPE. The bound seems to converge to
\be
\tilde{\Delta} = 2\Delta_\psi+1
\label{eq:bound}
\ee
as the number of derivatives is increased \cite{DefectBootstrap}. Figure \ref{fig:1DBound} shows a comparison of the numerical bound using 50 derivatives and the exact line \eqref{eq:bound}. The matching seems to deteriorate for higher $\Delta_\psi$. It is a well-known feature of numerical bootstrap using derivatives that convergence slows down dramatically as the external scaling dimension is increased, so one should not take this mismatch too seriously.

%%%%%%%%% Figure of the numerical 1D bootstrap bound
%%%%%%%%%

In fact, the bound can never be lower than $2\Delta_\psi+1$ because this value is saturated by the unitary solution to crossing corresponding to the generalized free real fermion in 1D, which also arises as the boundary dual of the free massive Majorana fermion in $\mathrm{AdS_2}$. The four-point function takes the form
\be
\langle \psi(x_1)\psi(x_2)\psi(x_3)\psi(x_4)\rangle = \frac{1}{|x_{12}|^{2\Delta_{\psi}}|x_{34}|^{2\Delta_{\psi}}}-\frac{1}{|x_{13}|^{2\Delta_{\psi}}|x_{24}|^{2\Delta_{\psi}}}
+\frac{1}{|x_{14}|^{2\Delta_{\psi}}|x_{23}|^{2\Delta_{\psi}}}\,,
\ee
where $\Delta_\psi$ can take an arbitrary positive value. In other words
\be
A(z) = 1 + \left(\frac{z}{1-z}\right)^{2\Delta_\psi}-z^{2\Delta_\psi}\,.
\ee
$A(z)$ can be decomposed in conformal blocks with positive coefficients, the spectrum being
\be
\Delta_j=2\Delta_{\psi}+2j+1\,,\quad j\in\mathbb{Z}_{\geq0}\,.
\label{eq:spectrum}
\ee
The primary operators appearing in this OPE are $\psi\!\overleftrightarrow{\partial}^{2j+1}\!\psi$, corresponding to two-particle states in $\textrm{AdS}_2$. The gap is indeed $2\Delta_\psi+1$. The existence of this solution together with evidence from the numerics suggests $2\Delta_\psi + 1$ is the optimal bootstrap bound. As explained in the previous subsection, proving this claim amounts to constructing (for each $\Delta_\psi > 0$) a nonzero functional $\omega_{\Delta_\psi}$  acting on functions $F_\Delta(z)$ defined in \eqref{eq:functionbasis}
\be
\omega_{\Delta_\psi}: F_\Delta(z)\mapsto \omega_{\Delta_\psi}(\Delta)\in\mathbb{R}
\ee
such that
\bee
&\omega_{\Delta_\psi}(0) = 0\\
&\omega_{\Delta_\psi}(\Delta) = 0\quad\textrm{for }\Delta=2\Delta_\psi + 2j+1\,,\quad j\in\mathbb{Z}_{\geq0}\\
&\omega_{\Delta_\psi}(\Delta)\geq 0\quad\textrm{for }\Delta\geq 2\Delta_\psi + 1
\label{eq:1DFunctionalProperties}
\eee
One of the main results of this paper is to construct such $\omega_{\Delta_\psi}$ explicitly when $\Delta_\psi$ is a positive integer or half-integer, and thus find the optimal bootstrap bound for these values.

%%%%%%%% Section: From derivative functionals towards the new basis
\section{From derivative functionals towards the new basis}\label{sec:commentsfunctional}
%%%%%%%%%% Subsection: Inadequacy of the z-derivatives
\subsection{Inadequacy of the $z$-derivatives and the Zhukovsky variable}\label{ssec:zhukovsky}
We will now discuss what the numerics have to say about the nature of the extremal functionals and introduce a new class of functionals that we use to construct the extremal functionals analytically in later sections. The discussion is framed in the context of 1D bootstrap but we expect analogous comments to apply in higher dimensions too.

\begin{figure}
\centering
\includegraphics[width=\textwidth]{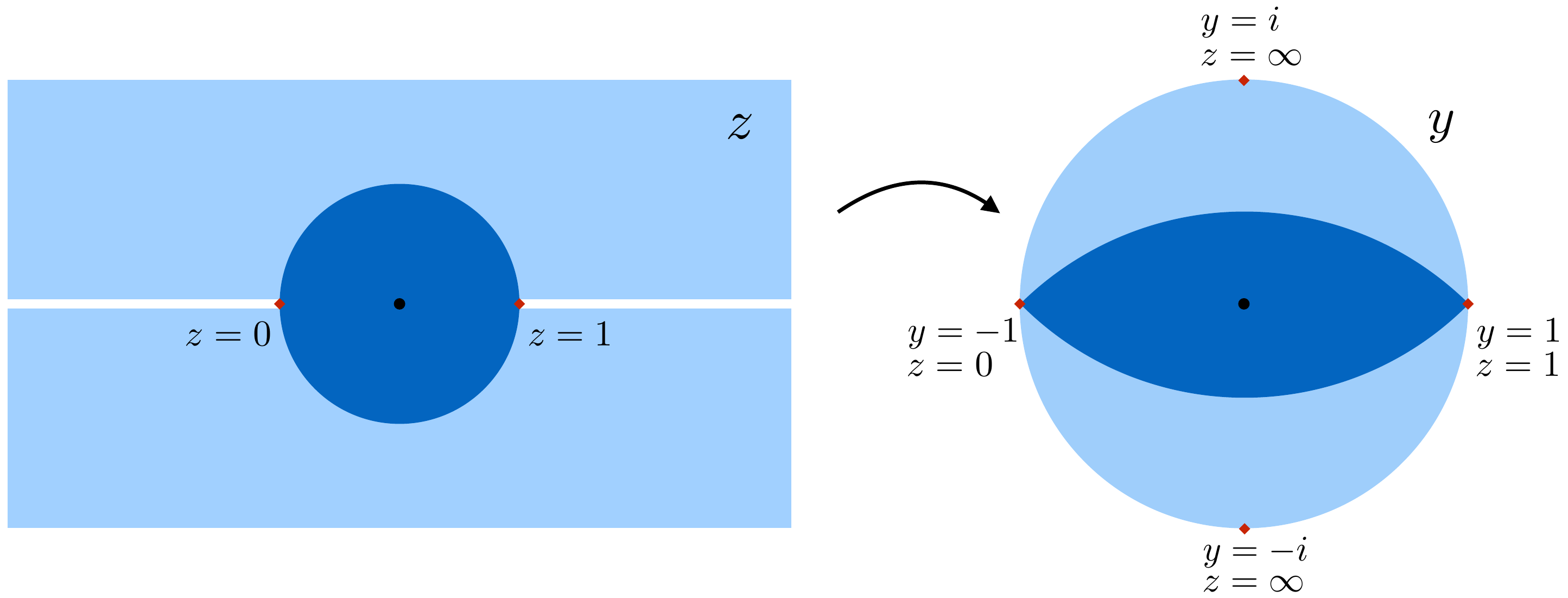}
\caption{The transformation \eqref{eq:ycoordinate} between the $z$ and $y$ coordinates. $z$-derivatives evaluated at $z=1/2$ can reconstruct the values of $F_{\Delta}(z)$ only in the dark-blue region, while $y$-derivatives at $y=0$ can reconstruct the values everywhere away from the branch cuts.}\label{fig:Zhukovsky}
\end{figure}

Numerical searches for functionals excluding candidate spectra have used the basis consisting of derivatives of functions $F_\Delta(z)$ defined through \eqref{eq:functionbasis}, evaluated at the crossing-symmetric point $z=1/2$. In practice, one truncates the space to derivatives of maximal degree $2N-1$. Let $\omega^{(N)}$ be the extremal functional in this truncated space and write
\be
\omega^{(N)} = \sum\limits_{j=1}^{N}\frac{a^{(N)}_{j}}{(2j-1)!}\left.\frac{d^{2j-1}}{dz^{2j-1}}\right|_{z=1/2}
\ee
with $a_j^{(N)}\in\mathbb{R}$. It is natural to wonder wether the extremal functional corresponding to the optimal bootstrap bound lies in the basis of derivatives, in other words whether $\omega^{(N)}$ converges in this basis as $N\rightarrow\infty$. At least for the 1D bootstrap problem at hand, the numerics indicate that this is not the case, and we expect the same happens in higher dimensions too. Since the functional is defined only up to an overall positive rescaling, let us normalize the leading coefficient as $|a_1^{(N)}|=1$. It turns out that (for any $\Delta_\psi$) as $N$ is increased, higher coefficients diverge as increasing powers of $N$
\be
a_{j}^{(N)}\overset{N\rightarrow\infty}{\sim} \beta_{j} N^{j-1}\,.
\label{eq:zdercoefs}
\ee
Hence the optimal extremal functional can not be a linear combination of derivatives at $z=1/2$. The result \eqref{eq:zdercoefs} resembles the evaluation of functions $F_\Delta(z)$ at a point that moves to infinity in the $z$-plane as $N$ increases. There is another instructive way to look at this divergence as follows. Equation \eqref{eq:be1D} holds everywhere in the complex $z$-plane away from the branch cuts located at $z\in(-\infty,0)$ and $z\in(1,\infty)$. However, derivatives at $z=1/2$ have access to information about $F_\Delta(z)$ only within the radius of convergence of $F_\Delta(z)$ around this point, i.e. only in $|z-1/2|<1/2$, see Figure \ref{fig:Zhukovsky}. The result \eqref{eq:zdercoefs} is thus telling us that the existence of the optimal bootstrap bound crucially relies on complex analytic behaviour of the functions $F_\Delta(z)$ outside of this disc.

There is a simple way to keep using derivatives at $z=1/2$ while getting access to the whole complex plane. We can map the complex plane without the two branch cuts to the interior of the unit disc via a version of the Zhukovsky transformation
\be
z(y) = \frac{(1+y)^2}{2(1+y^2)}\,,
\label{eq:ycoordinate}
\ee
illustrated in Figure \ref{fig:Zhukovsky}.

The points $z=0,1/2,1$ correspond to $y=-1,0,1$ respectively, while $z=\infty$ corresponds to the pair $y=\pm i$. The pair of branch cuts in the $z$-plane gets mapped to the unit circle. Crossing symmetry $z\leftrightarrow1-z$ gets mapped to $y\leftrightarrow-y$ and the Taylor expansion of $F_\Delta(z(y))$ around $y=0$ converges in the whole interior of the unit disc.
%%%%%%%%%
%%%%%%%%%

We can wonder whether the optimal extremal functional can be written as a linear combination of derivatives with respect to $y$ evaluated at $y=0$. For any finite $N$, the space of functionals generated by $\{\partial_z^{2j-1}|_{z=1/2},1\leq j\leq N\}$ and by $\{\partial_y^{2j-1}|_{y=0},1\leq j\leq N\}$ coincide. However, when we express the extremal functional for any finite $N$ in terms of the $y$-derivatives as
\be
\omega^{(N)} = \sum\limits_{j=1}^{N}\frac{b^{(N)}_{j}}{(2j-1)!}\left.\frac{d^{2j-1}}{dy^{2j-1}}\right|_{y=0}
\ee
and normalize $|b_1^{(N)}|=1$, we find the other coefficients converge
\be
b_{j}\equiv\lim_{N\rightarrow\infty} b_{j}^{(N)}\,,
\ee
in other words, the optimal extremal functional can be written as
\be
\omega = \sum\limits_{j=1}^{\infty}\frac{b_j}{(2j-1)!}\left.\frac{d^{2j-1}}{dy^{2j-1}}\right|_{y=0}
\ee
for some $b_j\in\mathbb{R}$. One could now try to fix coefficients $b_{j}$ leading to a functional with the desired properties \eqref{eq:1DFunctionalProperties}. However, it will turn out there is another representation of $\omega$ better suited for this task. This representation takes the form of an integral of the discontinuity of $F_\Delta(z)$ across the branch cut $z\in(1,\infty)$ against suitable integral kernels. To go from $y$-derivatives to such integrals, notice first that any derivative at $y=0$ of a function $f(y)$ holomorphic inside the unit disc can be written as a contour integral
\be
f^{(k)}(0) = \frac{k!}{2\pi i}\oint\limits_{\Gamma}\frac{\d y}{y} y^{-k}f(y)\,,
\ee
where the contour $\Gamma$ winds once around the origin and lies inside the unit disc. If $k\geq 1$, nothing changes with the insertion of an extra holomorphic term
\be
f^{(k)}(0) = \frac{k!}{2\pi i}\oint\limits_{\Gamma}\frac{\d y}{y} \left(y^{-k}-y^{k}\right)f(y)\,.
\ee
Taking $\Gamma$ to be the unit circle,\footnote{The contour can be taken all the way to the unit circle only in the absence of singularities of $f(y)$ on $|y|=1$. If these are present, we need to avoid them along infinitesimal arcs in the interior of the unit circle.} parametrized as $y=e^{i\theta}$, the integral becomes
\be
f^{(k)}(0) = \frac{k!}{i\pi}\int\limits_{0}^{2\pi}\!\d\theta\sin\left(k\theta\right)f\left(e^{i\theta}\right)\,.
\ee
Since we will be taking $f(y)=F_{\Delta}(z(y))$, we can assume $f(\bar{y}) = \bar{f(y)}$. Consequently, the last integral is only sensitive to the imaginary part of $f(y)$ on the unit circle. A general odd derivative functional can now be written as
\be
\omega(f) = \sum\limits_{j=1}^{\infty}\frac{b_{j}}{(2j-1)!}f^{(2j-1)}(0)
= \frac{1}{\pi}\int\limits_0^{2\pi}\!\d\theta\,g(\theta) \Im[f\left(e^{i\theta}\right)]\,,
\ee
where
\be
g(\theta) = \sum\limits_{j=1}^{\infty}b_{j}\sin\left[(2j-1)\theta\right]\,.
\label{eq:KernelFourier}
\ee
We can use symmetries of $F_\Delta(z)$ to simplify the result to
\be
\omega(F_\Delta) =\frac{4}{\pi}\int\limits_0^{\pi/2}\!\d\theta\,g(\theta) \Im[F_\Delta\left(z(e^{i\theta})\right)]
\label{eq:FunThetaKernel}
\ee
with $z(y)$ given by \eqref{eq:ycoordinate}. Since the unit circle in the $y$-coordinate corresponds to the pair of branch cuts of $F_\Delta$ in the $z$-coordinate, we see that we can write the functional as an integral of the imaginary part, or in other words discontinuity, of $F_\Delta(z)$ on the branch cut $z\in(1,\infty)$ against an appropriate integral kernel.\footnote{We thank Miguel Paulos for suggesting to look at functionals involving the discontinuity of the conformal block.} The coefficients $b_j$ are simply the Fourier coefficients of this kernel when the latter is written in the $\theta$ coordinate. However, there is a basis of functions on the branch cut which is more natural than the sines for the problem at hand. Namely the complete set of eigenfunctions of the conformal Casimir regular at the endpoints of the branch cut. In 1D, these eigenfunctions are simply Legendre polynomials of an appropriate coordinate. We are going to show soon how to fix the coefficients of the extremal functionals analytically in this basis. We can then always use the representation \eqref{eq:KernelFourier} and \eqref{eq:FunThetaKernel} to go back to the derivative basis.

%%%%%%%%%%% Subsection: The new basis
\subsection{The new basis}\label{ssec:newbasis}
\begin{figure}
\centering
\includegraphics[width=0.45\textwidth]{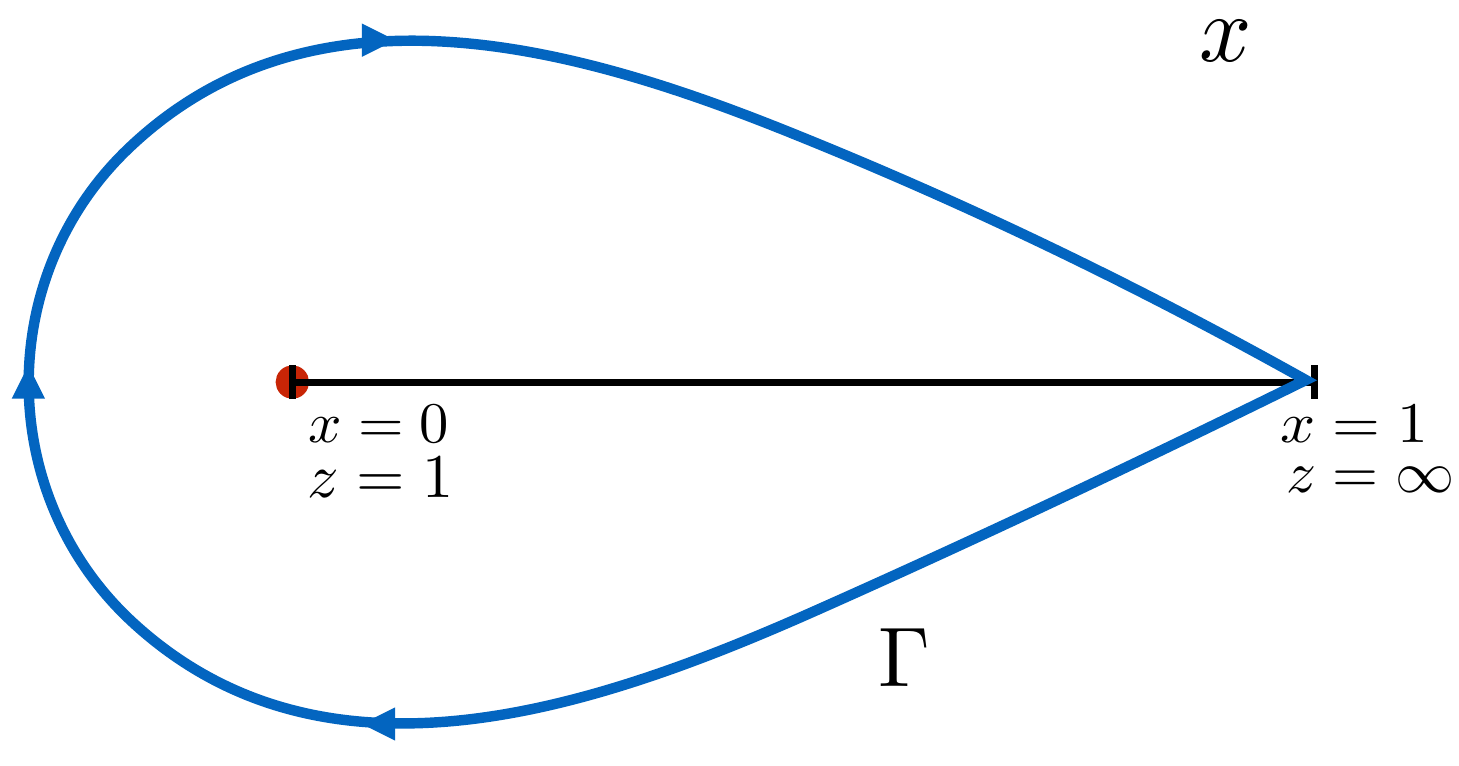}
\caption{The choice of integration contour leading to a well-defined action of $\omega$, equation \eqref{eq:functionalAction2}.}\label{fig:ContourX}
\end{figure}

As explained in the previous section, we want to write the extremal functionals as integral kernels applied to the imaginary part of $F_\Delta$ defined in \eqref{eq:functionbasis} on the branch cut $z\in(1,\infty)$.
Let us write
\be
F_\Delta(z) = g_\Delta(z)-g_\Delta(1-z)\,,
\ee
where
\bee
g_\Delta(z) &= z^{\Delta-2\Delta_\psi}{}_2F_1(\Delta,\Delta;2\Delta;z)\,.
\eee
It is convenient to map the branch cut to the unit interval $x\in (0,1)$ via
\be
x = \frac{z-1}{z}.
\ee
Let us denote
\bee
f^{+}_{\Delta}(x) &= \lim_{\epsilon\rightarrow 0^+}\frac{1}{\pi}\Im\left[g_{\Delta}(z(x)+ i \epsilon)\right]\\
f^{-}_{\Delta}(x) &= \lim_{\epsilon\rightarrow 0^+}\frac{1}{\pi}\Im\left[g_{\Delta}(1-z(x)- i \epsilon)\right]
\eee
It is not hard to evaluate $f_{\Delta}^{\pm}(x)$
\bee
f_\Delta^+(x)&= (1-x)^{2\Delta_\psi}\frac{\Gamma(2\Delta)}{\Gamma(\Delta)^2}{}_2F_1(\Delta,1-\Delta;1;x)\\
f_\Delta^-(x)&=- (1-x)^{2\Delta_\psi}\frac{\sin\left[\pi(\Delta-2\Delta_\psi)\right]}{\pi}x^{\Delta-2\Delta_\psi}{}_2F_1(\Delta,\Delta;2\Delta;x)\,.
\eee
$f_\Delta^+(x)$ is coming from the logarithmic branch cut of the direct channel conformal block starting at $z=1$. The crossed channel conformal block has a power-law singularity at $z=1$, so that $f_\Delta^-(x)$ is essentially the original conformal block with a sine prefactor. We are going to study functionals of the form
\bee
\omega(F_\Delta) &= 
\frac{1}{\pi}\int\limits_0^1\!\d x\,h(x)(1-x)^{-2\Delta_\psi}\Im\left[F_{\Delta}(z(x)+i\epsilon)\right]\\
&=\int\limits_0^1\!\d x\,h(x)(1-x)^{-2\Delta_\psi}\left[f_\Delta^+(x)-f_\Delta^-(x)\right]
\label{eq:functionalAction}
\eee
where $h(x)$ is a suitable integral kernel. We also explicitly eliminated the prefactor $(1-x)^{2\Delta_\psi}$ common to $f_\Delta^\pm(x)$. It is natural to expand $h(x)$ in the basis of solution of the conformal Casimir equation which are regular at $x=0,1$. Note that this choice breaks the symmetry between the direct and crossed conformal block since the Casimir equation is not invariant under $z\leftrightarrow 1-z$. However, we will see that the symmetry is partially restored by the full $h(x)$. The conformal Casimir equation in the direct channel written in the $x$-coordinate is just the Legendre differential equation
\be
\frac{\d }{\d x}\left[x(1-x)\frac{\d g(x)}{\d x}\right]+\Delta(\Delta-1)g(x)=0\,.
\label{eq:DifEq}
\ee
The solutions regular at $x=0,1$ have $\Delta=n\in\mathbb{N}$ and read
\be
p_n(x) = {}_2F_1(n,1-n;1;x)=(-1)^{n-1}P_{n-1}(2x-1)\,,
\ee
where $P_{m}(y)$ are the Legendre polynomials. $p_n(x)$ form a complete set of functions on $x\in[0,1]$ orthogonal with respect to the standard inner product with constant weight. We can expand $h(x)$ in this basis
\be
h(x) = \sum\limits_{n=1}^{\infty}a_n p_n(x)\,.
\label{eq:hExpansion}
\ee
In the following sections, we are going to present analytic formulas for $a_n$ that make $\omega$ into extremal functionals. Substituting \eqref{eq:hExpansion} into \eqref{eq:functionalAction}, the action of $\omega$ becomes (we use $\omega(F_\Delta)$ and $\omega(\Delta)$ interchangably in this paper)
\be
\omega(\Delta) = \sum\limits_{n=1}^\infty s(\Delta,n) a_n\,,
\label{eq:fActionS}
\ee
where
\be
s(\Delta,n)=s^+(\Delta,n)-s^-(\Delta,n)
\label{eq:s1D}
\ee
and we have defined overlaps of the imaginary part of $F_\Delta$ with our basis functionals
\be
s^{\pm}(\Delta,n) =  \int\limits_0^1\!\d x\,(1-x)^{-2\Delta_\psi}f_\Delta^{\pm}(x)p_n(x)\,.
\label{eq:sDefinition}
\ee
The overlaps can be found in a closed form as follows. $(1-x)^{-2\Delta_\psi}f^+_\Delta(x)$ satisfies the differential equation \eqref{eq:DifEq}, and so the overlap is particularly simple
\be
s^{+}(\Delta,n) =\frac{\Gamma(2\Delta)}{\Gamma(\Delta)^2}\frac{\sin[\pi(\Delta-n)]}{\pi (\Delta-n)(\Delta+n-1)}\,.
\label{eq:sPlus}
\ee
In particular, for $\Delta=m\in\mathbb{N}$, we find orthogonality
\be
s^+(m,n) = \frac{\Gamma(2m)}{\Gamma(m)^2}\int\limits_0^1\!\d x\, p_m(x)p_n(x) = \frac{\Gamma(2m-1)}{\Gamma(m)^2}\delta_{mn}\,.
\ee
The formula for $s^-(\Delta,n)$ is more complicated because the Casimir equations in the two channels do not coincide
\be
s^{-}(\Delta,n)
= (-1)^{n}\frac{\Gamma(2\Delta)}{\Gamma(\Delta)^2}\frac{\sin[\pi(\Delta-2\Delta_\psi)]}{\pi}R_{\Delta_\psi}(\Delta,n)\,,
\label{eq:sMinus}
\ee
where
\be
R_{\Delta_\psi}(\Delta,n)\equiv\frac{\Gamma(\beta)^2\Gamma(\gamma)^2}{\Gamma(\delta)\Gamma(\epsilon)\Gamma(\zeta)}
\setlength\arraycolsep{1pt}{}_4 F_3\left(\begin{matrix}&\beta& &\beta& &\gamma& &\gamma \\& &\delta& &\epsilon& &\zeta\end{matrix};1\right)
\label{eq:fdefinition}
\ee
with
\bee
\beta &= \Delta\\
\gamma &= \Delta-2\Delta_\psi + 1\\
\delta &= 2\Delta\\
\epsilon &= \Delta-2\Delta_\psi - n+2\\
\zeta &= \Delta -2\Delta_\psi + n+1\,.
\eee
A comment is in order concerning the regime of validity of \eqref{eq:sMinus}. $f_\Delta^-(x)p_n(x)=O(x^{\Delta-2\Delta_\psi})$ as $x\rightarrow 0$, so we would expect $s^-(\Delta,n)$ to be defined only for $\Delta>2\Delta_\psi - 1$. Indeed, $R_{\Delta_\psi}(\Delta,n)$ is an analytic function of $\Delta$ for $\Delta>2\Delta_\psi - 1$ with a simple pole at $\Delta = 2\Delta_\psi -1$. However, this pole is precisely cancelled by a zero of $\sin[\pi(\Delta - 2\Delta_\psi)]$ in the full expression for $s^-(\Delta,n)$. In fact, the formula \eqref{eq:sMinus} defines $s^-(\Delta,n)$ as a function analytic in $\Delta$ for any $\Delta\geq 0$. The reason is that the imaginary part of $F^-_\Delta$ on the branch cut is also the discontinuity of $F^-_\Delta$ across the branch cut. The integral \eqref{eq:sDefinition} can then be thought of as a contour integral in the complex $x$-plane with the contour starting at $x=1$, running under the branch cut, going around $x=0$ and coming back to $x=1$ above the branch cut. The contour can be deformed away from the real axis, and thus the singularity at $x=0$ is avoided, as illustrated in Figure \ref{fig:ContourX}, leading to a finite answer for any $\Delta\geq 0$. The proper generalization of our functionals \eqref{eq:functionalAction} to an arbitrary $\Delta$ is then
\be
\omega(F_\Delta) =\frac{1}{2\pi i} \int\limits_{\Gamma}\!\d x\,h(x)(1-x)^{-2\Delta_\psi}F_\Delta(z(x))\,.
\label{eq:functionalAction2}
\ee
Note that when passing from derivatives at $y=0$ to contour integrals as explained in the previous subsection, singularities of the integrand on the unit disc are avoided in the same manner. The bottom line is that the expression \eqref{eq:sMinus} can be trusted for any $\Delta\geq 0$. A further subtlety will later arise from the fact that $h(x)$, being an infinite linear combination of $p_n(x)$, develops a branch cut at $x\in(-\infty,0)$. Some care will then be needed to give meaning to \eqref{eq:functionalAction2}. However, no ambiguity is present when $h(x)$ is a single basis vector $p_n(x)$.
%%%%%%%%%
%%%%%%%%%

%%%%%%%%% Section: Constructing extremal functionals
\section{Constructing extremal functionals}\label{sec:extremalfunctionals}
%%%%%%%%%%% Subsection: General properties of the new basis
\subsection{General properties of the new basis}\label{ssec:properties}
Let us first explain why the new basis is particularly suitable for the construction of extremal functionals for 1D bootstrap. We will assume that $\Delta_\psi$ is an integer or half-integer, so that the oscillating factors in \eqref{eq:sPlus} and \eqref{eq:sMinus} are in phase. The action of the general functional \eqref{eq:functionalAction} can then be written as
\be
\omega(\Delta) = \frac{\Gamma(2\Delta)}{\Gamma(\Delta)^2}\frac{\sin(\pi\Delta)}{\pi}\sum\limits_{n=1}^\infty \hat{s}(\Delta,n)a_n\,,
\label{eq:functionalEv1}
\ee
where
\be
\hat{s}(\Delta,n) = \frac{(-1)^n}{(\Delta-n)(\Delta+n-1)}+(-1)^{2\Delta_\psi+n+1}R_{\Delta_\psi}(\Delta,n)\,.
\label{eq:sTilde}
\ee
In order to prove that the bootstrap bound is saturated by the generalized free fermion, we need to find coefficients $a_n$ so that $\omega(\Delta)$ has the properties listed in \eqref{eq:1DFunctionalProperties}. This means that $\Delta = 2\Delta_\psi + 1$ is a zero of odd order of $\omega(\Delta)$, while $\Delta = 2\Delta_\psi + 2j + 1$ is a zero of even order for any $j\in\mathbb{N}$. We will assume that in fact $\Delta = 2\Delta_\psi + 1$ is a simple zero and the higher $\Delta = 2\Delta_\psi + 2j + 1$ are all double zeros. The first key property of our basis is that it is in a sense a basis dual to the set of functions $F_\Delta$ with $\Delta\in\mathbb{N}$ and $\Delta\geq 2\Delta_\psi$. Indeed, suppose $\Delta = m\in\mathbb{N}$. The sine prefactor in \eqref{eq:functionalEv1} guarantees that $\omega(m)$ can only be non-zero if the sum in \eqref{eq:functionalEv1} has a pole at $\Delta=m$. As explained in the previous section, $R_{\Delta_\psi}(\Delta,n)$ has no poles for $\Delta > 2\Delta_\psi - 1$, so that in this range of $\Delta$ the pole must come from the first term in \eqref{eq:sTilde} and the summand $n=m$. We conclude that
\be
\omega(m) = \frac{\Gamma(2m-1)}{\Gamma(m)^2}a_m\quad\textrm{for }m\in\mathbb{N}\,,m\geq 2\Delta_\psi\,.
\label{eq:orthogonality}
\ee
In other words, the coefficient $a_m$ is proportional to the value of the functional at $\Delta = m$. It is also illuminating to consider the behaviour of the functionals at $\Delta = 0$. The prefactor in \eqref{eq:functionalEv1} has a double zero there, so any contribution must come from a double pole of $R_{\Delta_\psi}(\Delta,n)$. Indeed, there is such double pole, and its contribution can be written in a closed form
\be
\omega(0) = -\frac{1}{\Gamma(2\Delta_\psi)^2}\sum\limits_{n=2\Delta_\psi}^{\infty}(n-2\Delta_\psi+1)_{4\Delta_\psi - 2}\,a_n\,,
\label{eq:IdentityValue}
\ee
where $(a)_b$ is the Pochhammer symbol. This formula also follows from applying $\omega$ to the solution of crossing corresponding to the massive scalar in $\textrm{AdS}_2$ and using \eqref{eq:orthogonality}. Note that the Pochhammer symbol is only non-vanishing for $n\geq 2\Delta_\psi$ and that it is positive in that range. It follows that if $\omega(0) = 0$, as is required from an extremal functional, the coefficients $a_n$ for $n\geq 2\Delta_\psi$ must not all have the same sign. In particular, at least one of them, say $a_{n^*}$, is negative. Going back to \eqref{eq:orthogonality}, we conclude
\be
\omega(n^*) < 0\,.
\ee
Hence the bootstrap bound following from the existence of $\omega$ must be strictly above $\Delta = n^*$. The lowest choice is $n^* = 2\Delta_\psi$, and we conclude that any bound following from the functionals at hand must lie strictly above $\Delta = 2\Delta_\psi$. Of course, we already knew this thanks to the existence of the generalized free fermion solution, but it is reassuring to see it follow so naturally in the present language. Assuming that $\omega$ is an extremal functional with the spectrum of the generalized free fermion, we can now conclude from \eqref{eq:orthogonality} that
\bee
&a_{2\Delta_\psi} < 0\\
&a_{2\Delta_\psi + 2j-1} = 0\quad\textrm{for }j\in\mathbb{N}\\
&a_{2\Delta_\psi + 2j} >0\quad\textrm{for }j\in\mathbb{N}\,.
\label{eq:hFirstCond}
\eee
Moreover, the condition $\omega(0) = 0$ determines $a_{2\Delta_\psi}$ in terms of the higher $a_{2\Delta_\psi+2j}$ through \eqref{eq:IdentityValue} as
\be
a_{2\Delta_\psi} =- \sum\limits_{j=1}^{\infty}\frac{(2j+1)_{4\Delta_\psi-2}}{(4\Delta_\psi-2)!}a_{2\Delta_\psi + 2j}\,.
\ee
For the sum to be convergent, $a_n$ must decay at least as fast as
\be
a_n = O(n^{-\alpha})\quad\textrm{as }n\rightarrow\infty
\ee
with
\be
\alpha > 4\Delta_\psi - 1\,.
\ee
We will find out that in fact
\be
\alpha = 4\Delta_\psi +1\,.
\ee
In other words, the speed of convergence of the functionals to the optimal one improves with increasing $\Delta_\psi$ in the new basis. This is the exact opposite of what happens in the standard derivative basis, where high values of $\Delta_\psi$ require higher numbers of derivatives to achieve the same precision! This is one aspect of the particularly nice properties our functionals possess for $\Delta_\psi \gg 1$, elaborated on in section \ref{sec:ads}.

It remains to determine the values of $a_n$ for $1\leq n\leq 2\Delta_\psi - 1$ as well as $n=2\Delta_\psi + 2j$ with $j\in\mathbb{Z}_{\geq0}$. All these values of $a_n$ are fixed by requiring that $\Delta = 2\Delta_\psi + 2j - 1$ are double zeros of the functional \eqref{eq:functionalEv1} for $j\geq 2$, while $\Delta = 2\Delta_\psi + 1$ is a simple zero. The sine prefactor has a simple zero at all these locations, so the existence of a double zero implies the sum in \eqref{eq:functionalEv1} must itself vanish there. Denote
\be
\tilde{\omega}(\Delta) = \sum\limits_{n=1}^{\infty}\hat{s}(\Delta,n)a_n\,.
\ee
The conditions of $\omega(\Delta)$ having a simple zero and positive derivative at $\Delta = 2\Delta_\psi + 1$ and double zeros at higher $\Delta = 2\Delta_\psi + 2j - 1$ read
\be
\tilde{\omega}(2\Delta_\psi + 2j-1) = (-1)^{2\Delta_\psi+1}\delta_{j1}\quad\textrm{for }j\in\mathbb{N}\,,
\label{eq:AMatrixEq}
\ee
where the condition for $j=1$ fixes the arbitrary normalization of $\omega$. It is not hard to understand the mechanism of how these equations fix the values of non-zero $a_n$. We already know that $\tilde{\omega}(\Delta)$ has a simple pole at $\Delta= 2\Delta_\psi + 2j$ for $j\in\mathbb{N}$ 
\be
\tilde{\omega}(2\Delta_\psi+2j+\epsilon)\stackrel{\epsilon\rightarrow 0}{\sim} \frac{(-1)^{2\Delta_\psi}}{4\Delta_\psi+4j-1}\frac{a_{2\Delta_\psi+2j}}{\epsilon}\,.
\ee
Imagine changing $\Delta$ continuously from $\Delta = 2\Delta_\psi + 2j$ to $\Delta = 2\Delta_\psi+2j+2$. $\tilde{\omega}(\Delta)$ varies from plus to minus infinity, or vice versa, depending on the sign of $(-1)^{2\Delta_\psi}$. In any case, continuity implies
\be
\tilde{\omega}(\Delta)=0\quad\textrm{for some }\Delta\in(2\Delta_\psi+2j,2\Delta_\psi+2j+2)\quad\textrm{for all }j\in\mathbb{N}\,.
\ee
It is only for a specific choice of values of $a_n$ that all these zeroes of $\tilde{\omega}(\Delta)$ occur precisely at $\Delta = 2\Delta_\psi + 2j +1$. In order to find those values, it is useful to think of \eqref{eq:AMatrixEq} as an infinite matrix equation
\be
\sum\limits_{n=1}^{\infty}A_{j n}a_n = (-1)^{2\Delta_\psi + 1}\delta_{j1}\quad\textrm{for }j\in\mathbb{N}
\label{eq:MatrixEq}
\ee
with
\be
A_{j n} = \hat{s}(2\Delta_\psi+2j-1,n)\,.
\ee
If the linear map defined by matrix $A_{jn}$ were injective when acting on the subspace of $a_n$ that is not fixed to zero by conditions \eqref{eq:hFirstCond}, we could obtain $a_n$ simply as the first column of the inverse of $A_{jn}$
\be
a_n=(-1)^{2\Delta_\psi + 1} A^{-1}_{n 1}\,.
\label{eq:anFromInv}
\ee
The injectivity is in general violated for $\Delta_\psi\geq 3/2$, and we will address this subtlety in section \ref{sec:higherdelta}. Before we do that, let us first solve the case $\Delta_\psi=1/2$, where a closed formula for $h(x)$ can be found more directly.

\subsection{The extremal functional for $\Delta_\psi=1/2$}
\begin{figure}
\centering
\includegraphics[width=0.75\textwidth]{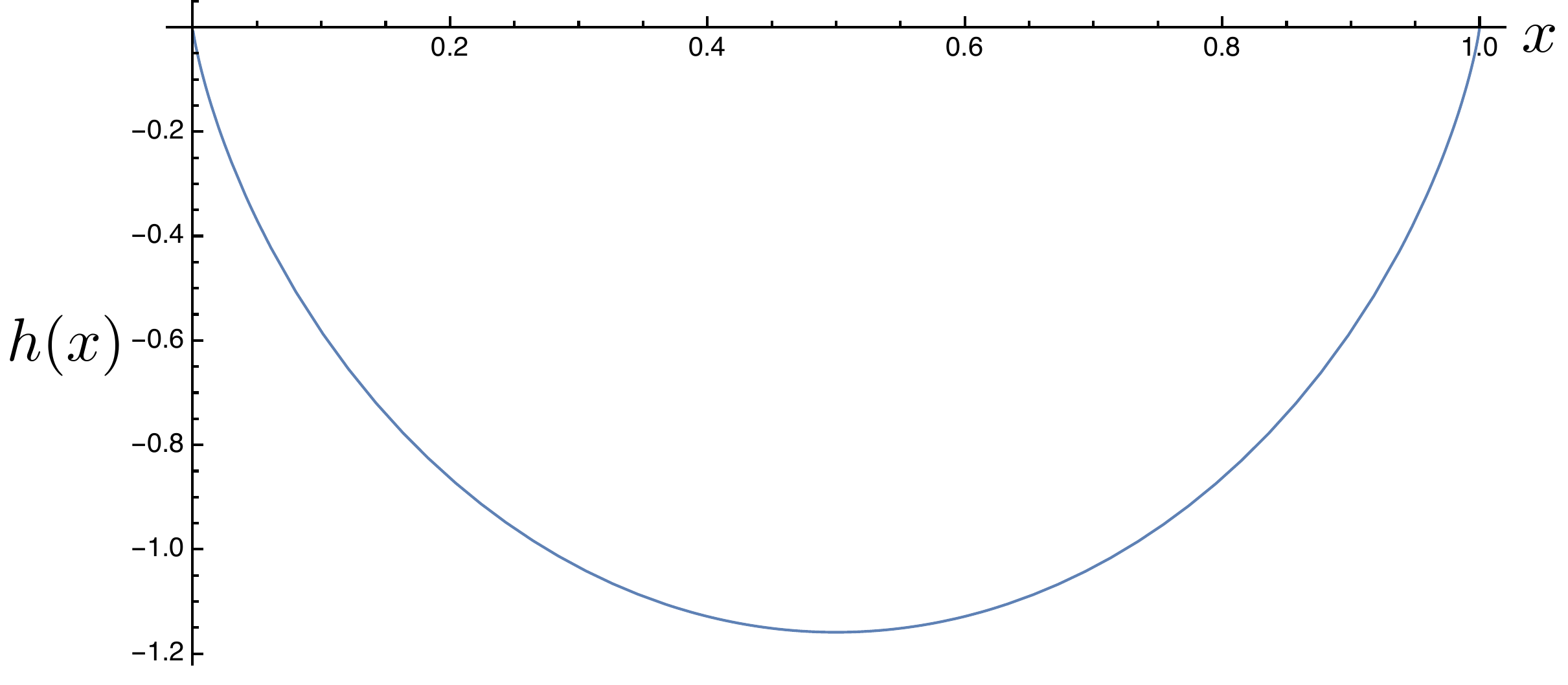}
\caption{The integral kernel $h(x)$ for $\Delta_\psi=1/2$, given by equation \eqref{eq:hXHalf}.}\label{fig:hXHalf}
\end{figure}

\begin{figure}
\centering
\includegraphics[width=0.8\textwidth]{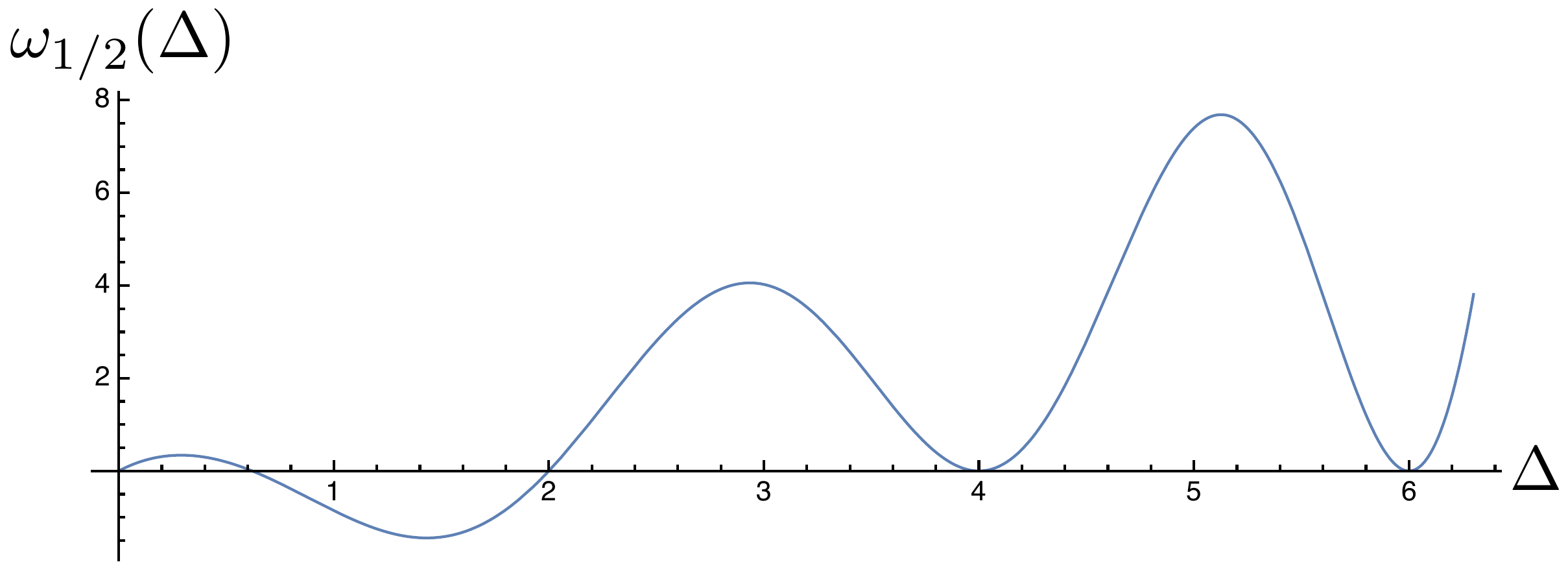}
\caption{The action of the extremal functional for $\Delta_\psi=1/2$, given by equation \eqref{eq:omegaHalfAction}.}\label{fig:omegaHalf}
\end{figure}

Notice first that when $\Delta_\psi=1/2$, orthogonality \eqref{eq:orthogonality} is valid for all $m\in\mathbb{N}$. Since the extremal functional corresponding to the generalized free fermion should vanish for all $\Delta\in2\mathbb{N}$, we conclude that only $a_{n}$ with $n$ odd can be nonzero
\be
h(x) =\sum_{n\in2\mathbb{N}-1}a_{n}p_{n}(x)\,.
\label{eq:hHalfSum}
\ee
It follows from the symmetry property of the basis functions
\be
p_n(x) = (-1)^{n-1}p_{n}(1-x)
\ee
that $h(x) = h(1-x)$. We would now like to impose the conditions on derivatives \eqref{eq:AMatrixEq}
\bee
\omega'(2)&>0\\
\omega'(2j+2) &= 0\quad\textrm{for }j\in\mathbb{N}\,,
\eee
which take the explicit form
\be
\sum_{n\in2\mathbb{N}-1}
\left[-\frac{1}{(\Delta-n)(\Delta+n-1)}-R_{\frac{1}{2}}(\Delta,n)\right]a_{n} =
\begin{dcases}
1\quad&\textrm{for }\Delta = 2\\
0\quad&\textrm{for }\Delta = 2j +2\,,j\in\mathbb{N}\,,
\end{dcases}
\label{eq:DersHalf}
\ee
with $R_{\Delta_\psi}(\Delta,n)$ defined in \eqref{eq:fdefinition}. Rather than solving these equations directly for $a_n$, we will first express them in terms of scalar products of functions on the unit interval. Let us first define the following functions for $\Delta\in2\mathbb{N}$
\bee
q_{\Delta}(x) &= Q_{\Delta-1}(2x-1)\\
r_{\Delta}(x) &= \frac{\Gamma(\Delta)^2}{2\Gamma(2\Delta)}\left[x^{\Delta-1}{}_2F_1(\Delta,\Delta;2\Delta;x)+ (1-x)^{\Delta-1}{}_2F_1(\Delta,\Delta;2\Delta;1-x)\right]\\
s_{\Delta}(x) &= q_\Delta(x) - r_\Delta(x)\,,
\eee
where $Q_m(y)$ is the Legendre function of the second kind. When $\Delta\in2\mathbb{N}$ both $q_\Delta(x)$ and $r_\Delta(x)$ are symmetric under $x\leftrightarrow 1-x$ and hence
\be
s_\Delta(1-x) = s_\Delta(x)\,.
\ee
The leading logarithmic divergence and constant term of $q_\Delta(x)$ and $r_\Delta(x)$ at the boundary of the interval precisely cancel and we find $s_\Delta(0) = s_\Delta(1) = 0$. Define the usual scalar product on the space of function on the unit interval
\be
\left\langle f,g\right\rangle = \int\limits_0^1\!\d x f(x)g(x)\,.
\ee
Unlike the Legendre polynomials, $q_\Delta(x)$ are not orthogonal with respect to this scalar product. However, the corrected functions $s_\Delta(x)$ are mutually orthogonal
\be
\langle s_{\Delta},s_{\Delta'}\rangle = \frac{\pi^2}{4(2\Delta-1)} \delta_{\Delta \Delta'}\quad\textrm{for }\Delta,\Delta'\in2\mathbb{N}\,.
\ee
Indeed, $s_\Delta(x)$ form an orthogonal basis for functions on $x\in(0,1)$ satisfying $f(x) = f(1-x)$. The crucial observation arises from computing the scalar product of $s_{\Delta}(x)$ and $p_n(x)$ with $n$ odd
\be
\langle s_{\Delta},p_n\rangle = -\frac{1}{(\Delta-n)(\Delta+n-1)}-R_{\frac{1}{2}}(\Delta,n)\,.
\ee
The first and second term come from the overlap with $q_\Delta(x)$ and $r_\Delta(x)$ respectively. We recognize that the scalar product is precisely the coefficient with which $p_n$ contributes to the derivative of $\omega(\Delta)$! Equation \eqref{eq:DersHalf} can thus be written simply as
\be
\sum\limits_{n=1}^{\infty}\langle s_{2j},p_n\rangle a_n = \delta_{j1}\quad\textrm{for }j\in\mathbb{N}\,.
\ee
In other words
\be
\langle s_{2j},h\rangle = \delta_{j1}\quad\textrm{for }j\in\mathbb{N}\,,
\ee
where $h(x)$ is the sought integral kernel. Since $s_{2j}(x)$ are orthogonal, the last equation is telling us precisely that $h(x)$ must be proportional to $s_2(x)$. Hence, up to an overall irrelevant positive constant
\be
h(x) = s_2(x) =
\frac{1}{x(1-x)}-1+
\left[\frac{x \left(2 x^2-5 x+5\right)}{2 (1-x)^2} \log (x)+(x\leftrightarrow 1-x)\right]\,.
\label{eq:hXHalf}
\ee
Figure \ref{fig:hXHalf} shows the shape of $h(x)$. Note that $h(0)=h(1)=0$, so that the integral in \eqref{eq:functionalAction} is convergent on both ends.
%%%%%%%%%
%%%%%%%%%
It is also possible to find a closed formula for the coefficients $a_n$. Define the function
\be
\Omega_{\frac{1}{2}}(\Delta) =
\frac{1}{(\Delta-2)(\Delta+1)}
-\left[\Delta(\Delta-1)+\frac{1}{2}\right]\Psi'\left(\frac{\Delta}{2}\right)-2\,,
\label{eq:htilden}
\ee
where
\be
\Psi(z) =\frac{\d}{\d z}\log\left[\frac{\Gamma\left(z+\frac{1}{2}\right)}{\Gamma\left(z\right)}\right] = \psi(z+1/2)-\psi(z)\,,
\label{eq:PsiFun}
\ee
with $\psi(z)$ the digamma function. Coefficients $a_n$ take the form
\be
a_n =
\begin{dcases}
(2n-1)\Omega_{\frac{1}{2}}(n)\quad&\textrm{for }n\textrm{ odd}\\
0\quad&\textrm{for }n\textrm{ even}\,.
\end{dcases}
\ee
The nonzero $a_n$ decay like $a_n=O(n^{-3})$ as $n\rightarrow\infty$, a special case of the general formula $a_n=O(n^{-4\Delta_\psi - 1})$. It is not hard to find a formula for the action of the extremal functional on $F_\Delta$ for any $\Delta$. It follows from \eqref{eq:orthogonality} that
\be
\omega_{\frac{1}{2}}(\Delta) =
\begin{dcases}
\frac{\Gamma(2\Delta)}{\Gamma(\Delta)^2}\Omega_{\frac{1}{2}}(\Delta)\quad&\textrm{for }\Delta\in\mathbb{N}\textrm{ odd}\\
0\quad&\textrm{for }\Delta\in\mathbb{N}\textrm{ even}\,.
\end{dcases}
\ee
The simplest meromorphic function of $\Delta$ interpolating between these values is
\be
\omega_{\frac{1}{2}}(\Delta) = \frac{\Gamma(2\Delta)}{\Gamma(\Delta)^2}\sin^2\left(\frac{\pi\Delta}{2}\right)\Omega_{\frac{1}{2}}(\Delta)\,,
\label{eq:omegaHalfAction}
\ee
which turns out to be the correct formula. The function $\omega_{\frac{1}{2}}(\Delta)$ is plotted in Figure \ref{fig:omegaHalf}. As expected, $\omega_{\frac{1}{2}}(\Delta)$ has double zeros at $\Delta = 2+2j$, $j\in\mathbb{N}$. The simple pole of $\Omega_{\frac{1}{2}}(\Delta)$ at $\Delta=2$ makes this into a simple zero of $\omega_{\frac{1}{2}}(\Delta)$.
%%%%%%%%%
%%%%%%%%%

\begin{figure}
\centering
\includegraphics[width=0.8\textwidth]{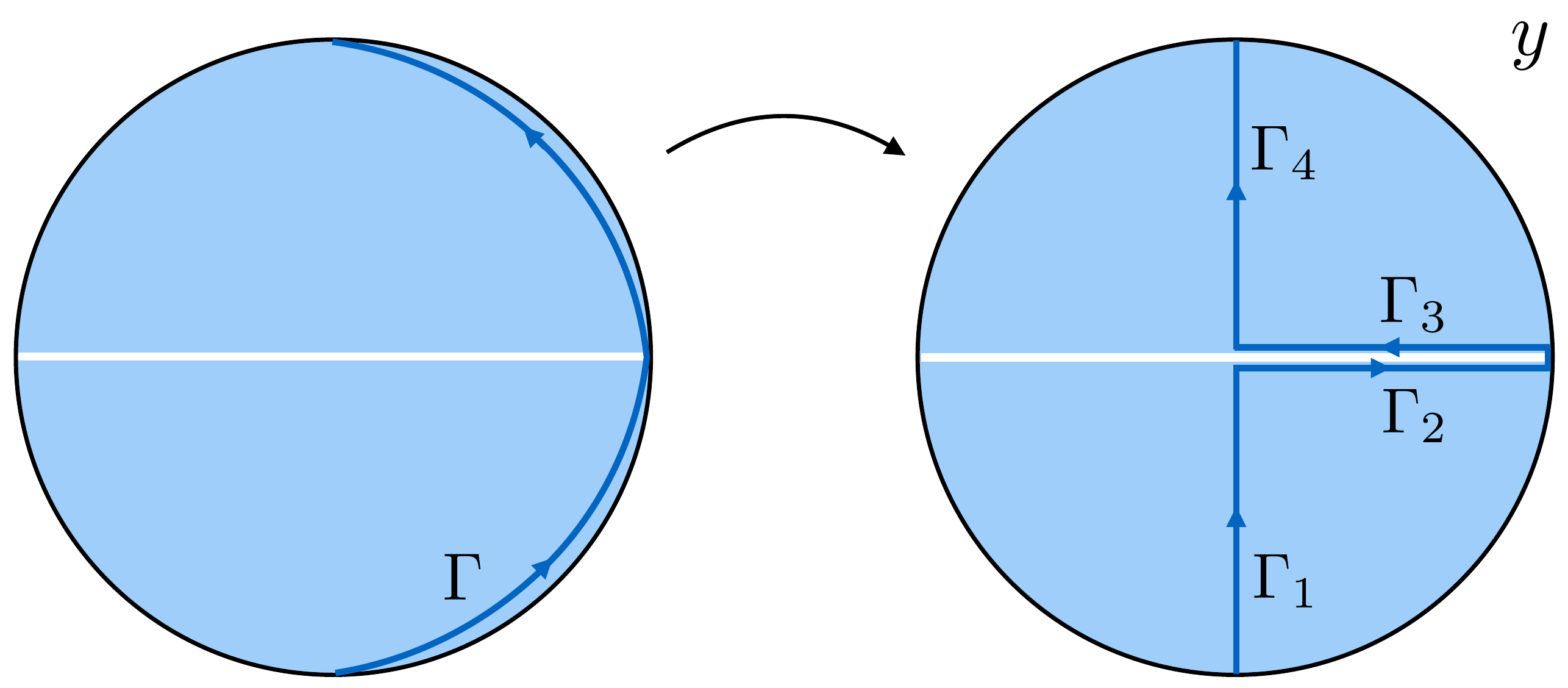}
\caption{The contour deformation in the $y$-coordinate used to find a closed formula \eqref{eq:bjHalfExact} for the coefficients $b_j$ of the extremal functional in the $y$-derivative basis. The white stripe shows the branch cut of $h(x)$.}\label{fig:contourDef}
\end{figure}

\begin{figure}
\centering
\includegraphics[width=0.7\textwidth]{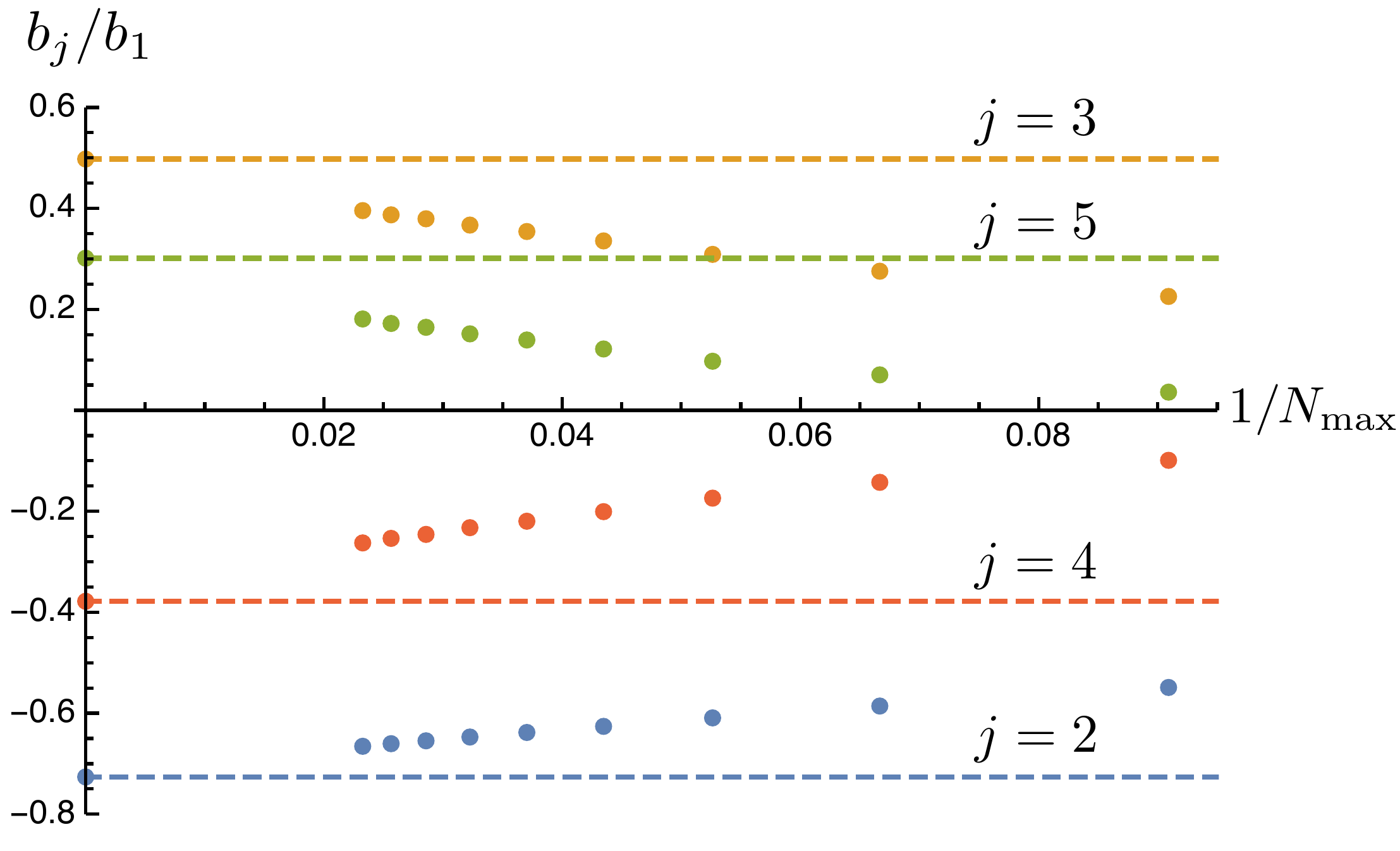}
\caption{Comparison of the analytic extremal functional in the derivative basis \eqref{eq:bjHalfExact} (dashed lines) and the numerical bootstrap extremal functionals (dots) for $\Delta_\psi=1/2$. $b_j$ is the coefficient of the $y$-derivative of order $2j-1$ and $N_{\textrm{max}}$ is the order of the maximal $z$-derivative used in the numerics.}\label{fig:yDerCoefs0p5}
\end{figure}

We can use the explicit formula for $h(x)$ to produce a closed formula for the coefficients $b_j$ of the functional in the basis of $y$-derivatives evaluated at $y=0$
\be
\omega = \sum\limits_{j=1}^{\infty}\frac{b_j}{(2j-1)!}\left.\frac{d^{2j-1}}{dy^{2j-1}}\right|_{y=0}\,,
\ee
described in subsection \ref{ssec:zhukovsky}. Recall from \eqref{eq:functionalAction2} that the action of the functional is\footnote{By a slight abuse of notation, we write $F_\Delta(x)$, $F_\Delta(y)$ instead of $F_\Delta(z(x))$, $F_\Delta(z(y))$ here and in the following.}
\be
\omega(\Delta) = \frac{1}{2\pi i}\int\limits_{\Gamma}\!\d x \frac{h(x)}{1-x}F_\Delta(x)\,,
\label{eq:omegaHalfCont}
\ee
where the contour $\Gamma$ is shown in Figure \ref{fig:ContourX}. However, formula \eqref{eq:hXHalf} shows that $h(x)$ has a branch cut on $x\in(-\infty,0)$ so the contour integral seems not well-defined since its value depends on where the contour $\Gamma$ intersects the branch cut of $h(x)$. It is possible to see that the choice of $\Gamma$ that reproduces the correct action \eqref{eq:functionalEv1} is the one intersecting the real axis arbitrarily close to $x=0$. In other words, we recover the prescription \eqref{eq:functionalAction}. The advantage of the description using a contour integral passing arbitrarily close to $x=0$ as opposed to \eqref{eq:functionalAction} is that the former will be valid for any $\Delta_\psi$. To pass to the derivative basis, let us start by transforming the integral \eqref{eq:omegaHalfCont} to the Zhukovsky coordinate $y$ defined by \eqref{eq:ycoordinate}, which is related to $x$ via
\be
x(y) = - \left(\frac{y-1}{y+1}\right)^2\,.
\ee
The contour $\Gamma$ gets mapped to the blue curve in the left half of Figure \ref{fig:contourDef}.
%%%%%%%%%
%%%%%%%%%
$F_\Delta(y)$ is holomorphic inside the unit circle of variable $y$. $h(x(y))$ has a branch cut along $y\in(-1,1)$ coming from the $\log(x)$ term in \eqref{eq:hXHalf}. We can deform the contour as in Figure \ref{fig:contourDef} to get a contour with four components as follows:
\bee
\Gamma_1:\quad &y(t)=-i + i t\quad &t\in[0,1]\\
\Gamma_2:\quad &y(t)= t-i\epsilon\quad &t\in[0,1]\\
\Gamma_3:\quad &y(t)=1- t+i\epsilon\quad &t\in[0,1]\\
\Gamma_4:\quad &y(t)= i t\quad &t\in[0,1]
\eee
where $\epsilon\rightarrow 0^+$. The integrals along $\Gamma_1$ and $\Gamma_4$ combine to depend only on the imaginary part of $h(x(y))$ along the imaginary axis, the result being
\be
 \frac{1}{2\pi i}\int\limits_{\Gamma_1\cup\Gamma_4}\!\!\!\d x \frac{h(x)}{1-x}F_\Delta(x) = 
 -\int\limits_{\Gamma_{4}}\!\d y \frac{3 y^4+26 y^2+3}{\left(1-y^2\right)^3}F_{\Delta}(y)\,.
 \label{eq:gamma14}
\ee
The integrals along $\Gamma_2$ and $\Gamma_3$ combine to depend only on the discontinuity of $h(x(y))$ across the branch cut, the result being
\bee
 &\frac{1}{2\pi i}\int\limits_{\Gamma_2\cup\Gamma_3}\!\!\!\d x \frac{h(x)}{1-x}F_\Delta(x) = \\
 &=\int\limits_{0}^1\!\d y\frac{(1-y)^3 \left(3 y^4+3 y^3+8 y^2+3 y+3\right)}{(y+1)^3 \left(y^2+1\right)^3}F_{\Delta}(y)\,.
 \label{eq:gamma23}
\eee
The coefficients $b_j$ can now be found by substituting the Taylor expansion of $F_\Delta(y)$ around $y=0$ into \eqref{eq:gamma14} and \eqref{eq:gamma23}. We find
\bee
b^{(1)}_j &= -\int\limits_{\Gamma_{4}}\!\d y \frac{3 y^4+26 y^2+3}{\left(1-y^2\right)^3} y^{2j-1} =\\
&=\frac{(-1)^j}{4} \left[8 (2 j-1)-\left(4 j-1\right) \left(4 j-3\right)\Psi\left(\frac{j}{2}\right)\right]\,,
\eee
with $\Psi(z)$ defined in \eqref{eq:PsiFun}. Similarly,
\bee
b^{(2)}_j &=
\int\limits_{0}^1\!\d y\frac{(1-y)^3 \left(3 y^4+3 y^3+8 y^2+3 y+3\right)}{(y+1)^3 \left(y^2+1\right)^3}y^{2j-1}=\\
&=
\frac{(4 j-3) (4 j-1)}{2}\Psi (j)-\frac{(2 j-3) (2 j+1)}{16} \Psi\left(\frac{2 j+1}{4}\right)-\frac{15 j-11}{4}
\,.
\eee
The derivative coefficients are simply the sum
\be
b_j = b_j^{(1)} + b_{j}^{(2)}\,.
\label{eq:bjHalfExact}
\ee
The first few values of $b_j$ read
\bee
b_1 &= \frac{3}{4}-\frac{3}{16}\pi -\frac{3}{2} \log (2)\approx -0.878769\\
b_2 &=\frac{55}{4}-\frac{5}{16}\pi -\frac{35}{2} \log (2)\approx 0.638177\\
b_3 &=\frac{119}{4}+\frac{21}{16}\pi-\frac{99}{2} \log (2)\approx -0.437445\\
b_4 &=\frac{307}{4}-\frac{45}{16} \pi -\frac{195}{2} \log (2)\approx0.332421\\
&\;\; \vdots
\eee
Figure \ref{fig:yDerCoefs0p5} shows a comparison between the exact values for $b_j$ and those obtained by standard numerical bootstrap when derivatives are truncated to maximal degree $N_{\textrm{max}}$. Only ratios of derivative coefficients can be compared since the overall normalization is arbitrary. The dashed lines are obtained from the exact values \eqref{eq:bjHalfExact} while dots of the same color correspond to the appropriate numerical bootstrap results. The plot shows convincing evidence that the numerical bootstrap tends to the exact answer as $N_\textrm{max}\rightarrow\infty$.

%%%%%%%%%
%%%%%%%%%

Finally, we would like to point out that although our choice of basis for the bootstrap functionals breaks the symmetry between direct and crossed channels $z\leftrightarrow 1-z$, the full extremal functional enjoys a version of this symmetry. $z\leftrightarrow 1-z$ corresponds to $x\leftrightarrow 1/x$, so such symmetry can only be a property of the full sum \eqref{eq:hHalfSum}. The integral kernel can be rewritten
\be
\frac{h(x)}{1-x}\d x = \tilde{h}(z)\d z\,,
\ee
with
\be
\tilde{h}(z)=
1-\frac{1}{z(1-z)}-
\left[\frac{(1-z) \left(2 z^2+z+2\right)}{2 z^2} \log (z-1)+
\frac{z \left(2z^2-5z+5\right)}{2 (1-z)^2} \log (z)
\right]\,.
\ee
The last expression is symmetric under $z\leftrightarrow1-z$ up to a minus sign picked up by the argument of the first logarithm. A similar property holds also for higher $\Delta_\psi$, see Appendix \ref{app:closed}. It would be interesting to see if the same symmetry is present for arbitrary $\Delta_\psi$.

%%%%%%%%%%%%%%%%
\section{Higher values of $\Delta_\psi$}\label{sec:higherdelta}
%%%%%%%%%%%%%%%%%%%%%%
\subsection{Linear dependence of elementary functionals}\label{ssec:lindep}
Before we write down analytic formulas for the extremal functionals when $\Delta_\psi$ is an arbitrary positive integer or half-integer, we need to explain one subtlety. The functions $p_n(x)$, $n\in\mathbb{N}$ are linearly independent so we would expect that their overlaps with $F_\Delta$ given by
\be
s(\Delta,n) = \frac{\Gamma(2\Delta)}{\Gamma(\Delta)^2}\frac{\sin(\pi\Delta)}{\pi}
\left[\frac{(-1)^n}{(\Delta-n)(\Delta+n-1)}+(-1)^{2\Delta_\psi+n+1}R_{\Delta_\psi}(\Delta,n)\right]
\ee
are linearly independent as functions of $\Delta$. This is easily seen to be true for $n\geq 2\Delta_\psi$ thanks to the orthogonality \eqref{eq:orthogonality}. However, it is generally not true for $1\leq n\leq2\Delta_\psi - 1$, in spite of the fact that all $s^+(\Delta,n)$ as well as all $s^-(\Delta,n)$ are linearly independent. In other words, the linear dependence arises thanks to precise cancellations between the contribution of the direct and crossed conformal blocks. A direct computation leads to the following examples for small $\Delta_\psi$
\bee
\Delta_\psi = 1:\quad &s(\Delta,1) = 0\\
\Delta_\psi = \frac{3}{2}:\quad &s(\Delta,1) = -s(\Delta,2)\\
\Delta_\psi = 2:\quad & s(\Delta,1) = s(\Delta,2) = -\frac{1}{5}s(\Delta,3)
\eee
where the equalities hold for arbitrary $\Delta\in\mathbb{R}$. It is natural to ask what is the kernel of the map
\be
\varphi:h(x)\mapsto  \int\limits_{\Gamma}\!\d x\,h(x)(1-x)^{-2\Delta_\psi}F_\Delta(x)\,.
\ee
Since $p_n(x)$ is a polynomial of degree $n-1$, the kernel lies within the space of polynomials of degree at most $2\Delta_\psi - 2$. In fact, it is possible to give a simple explicit description of $\ker\varphi$ as follows
\be
\ker\varphi=
\begin{dcases}
\left\langle x^a(x-1)^{2b},a+b=\Delta_\psi - 1\right\rangle\quad&\textrm{for }\Delta_\psi\in\mathbb{N}\\
\left\langle x^a(x-1)^{2b+1},a+b=\Delta_\psi - 3/2\right\rangle\quad&\textrm{for }\Delta_\psi\in\mathbb{N}-\frac{1}{2}\,,
\end{dcases}
\label{eq:ker}
\ee
where $\langle \alpha\rangle$ denotes the span of the set $\alpha$ and $a,b\in\mathbb{Z}_{\geq0}$. We see that $s(\Delta,n)$ with $1\leq n\leq 2\Delta_\psi -1$ considered as functions of $\Delta$ generate a space of roughly half the full dimensionality, specifically
\be
\dim\left(\left\langle s(\Delta,n),1\leq n \leq 2\Delta_\psi - 1\right\rangle\right)=
\begin{dcases}
\Delta_\psi - 1\quad\textrm{for }\Delta_\psi\in\mathbb{N}\\
\Delta_\psi - \frac{1}{2}\quad\textrm{for }\Delta_\psi\in\mathbb{N}-\frac{1}{2}\,.
\end{dcases}
\ee
In fact, in both cases $\left\langle s(\Delta,n),1\leq n \leq 2\Delta_\psi - 1\right\rangle$ is generated by the linearly independent set $\{s(\Delta,2j),1\leq j\leq \lfloor\Delta_\psi-1/2\rfloor\}$.

The linear dependence implies that some columns of matrix $A_{jn}$ appearing \eqref{eq:MatrixEq} are linearly dependent and we can not find $a_n$ simply by inverting the full $A_{jn}$. However, $A_{jn}$ can be inverted when $n$ is restricted to a set corresponding to linearly independent functions $s(\Delta,n)$. Any two solutions of \eqref{eq:MatrixEq} differ by a vector $\delta a_n$ corresponding to a function in $\ker\varphi$
\be
\sum\limits_{n=1}^{2\Delta_\psi - 1}\delta a_np_{n}(x)\in\ker\varphi\,.
\ee
We will see that this redundancy can be eventually fixed by requiring that the integral kernel has a Fourier expansion as in \eqref{eq:KernelFourier}, in other words that $(1-x)^{-2\Delta_\psi}h(x)$ has at most a logarithmic singularity at $x=1$.

%%%%%%%%%%%%%%%%%%%%%%%%
\subsection{Extremal functionals for $\Delta_\psi\in\mathbb{N}$}\label{ssec:FunctionalsInt}
We are now ready to write down an explicit formula for $a_n$ leading to the extremal functional $\omega_{\Delta_\psi}$ corresponding to the optimal bootstrap bound $2\Delta_\psi + 1$ for $\Delta_\psi\in\mathbb{N}$. Recall from subsection \ref{ssec:properties} that $a_{2\Delta_\psi + 2j -1} = 0$ for $j\in\mathbb{N}$ and from subsection \ref{ssec:lindep} that it is sufficient to keep only even $n$ from among $1\leq n\leq 2\Delta_\psi-1$. Therefore, $h(x)$ can be written as
\be
h(x) = \sum_{n\in2\mathbb{N}}a_{n}p_{n}(x)\,.
\label{eq:hXInteger}
\ee
Consequently, $h(1-x) = -h(x)$ since $p_n(1-x) = (-1)^{n-1}p_n(x)$. Note that the map $x\leftrightarrow 1-x$ corresponds to $z\leftrightarrow z/(z-1)$ and hence to swapping positions $x_3$ and $x_4$ in the four-point function. It would be interesting to see if there is a physical interpretation of this symmetry of $h(x)$.

$a_{2k}$ satisfy the equation \eqref{eq:MatrixEq}
\be
\sum\limits_{k=1}^\infty \tilde{A}_{jk} a_{2k} = - c_{\Delta_\psi}\delta_{j1}\quad\textrm{for } j\in\mathbb{N}\,,
\label{eq:MatrixEq2}
\ee
where
\be
\tilde{A}_{jk} = \hat{s}(2\Delta_\psi+2j-1,2k)
\ee
with $\hat{s}(\Delta,n)$ given by \eqref{eq:sTilde} and $c_{\Delta_\psi}$ is an arbitrary positive normalization. $\tilde{A}_{jk}$ is now non-singular when the $j,k$ indices are truncated to an arbitrary range $j,k\in\{1,\ldots J\}$. In spite of the rather complicated form of the entries of $\tilde{A}_{jk}$, the normalizable solution of equation \eqref{eq:MatrixEq2} can be written in a closed form for arbitrary $\Delta_\psi$ as follows. Define
\bee
\alpha_{\Delta_\psi}(\Delta,m) &=
\left[1+4 m \left(\Delta _{\psi }- m\right)\right]
\frac{\Gamma (4 m+1)\Gamma\left(m-\frac{1}{2}\right)^2
\Gamma \left(\frac{\Delta+1}{2}\right)^2
\Gamma\left(\frac{\Delta+1}{2}-m\right)^2}
{2^{8 m}\pi(4m-1)\Gamma (m+1)^2
\Gamma \left(\frac{\Delta}{2}\right)^2
\Gamma \left(\frac{\Delta}{2}+m\right)^2}\\
\beta_{\Delta_\psi}(\Delta,m) &=
\left[1-2\left(\Delta_{\psi }-m\right)\right]
\frac{ \Gamma (4 m+1)\Gamma\left(m+\frac{1}{2}\right)^2
\Gamma \left(\frac{\Delta+1}{2}\right)^2
\Gamma\left(\frac{\Delta-1}{2}-m\right)\Gamma\left(\frac{\Delta+1}{2}-m\right)}
{2^{8 m+1}\pi\Gamma (m+1)^2
\Gamma \left(\frac{\Delta}{2}\right)^2
\Gamma \left(\frac{\Delta}{2}+m\right)\Gamma \left(\frac{\Delta}{2}+m+1\right)}\,.
\eee
Use these to define $\Omega_{\Delta_\psi}(\Delta)$ for $\Delta_\psi\in\mathbb{N}$
\be
\Omega_{\Delta_\psi}(\Delta)=
\Delta(\Delta-1)+\Delta_\psi+
\sum\limits_{m=0}^{\Delta_\psi}\left[\alpha_{\Delta_\psi}(\Delta,m) +\beta_{\Delta_\psi}(\Delta,m) \right]\,.
\label{eq:OmegaInteger}
\ee
The formula for $a_n$ is then
\be
a_n =
\begin{dcases}
(2n-1)\Omega_{\Delta_\psi}(n)\quad&\textrm{for }n\textrm{ even}\\
0\quad&\textrm{for }n\textrm{ odd}
\end{dcases}
\label{eq:hInteger}
\ee
It can be checked that $a_{2\Delta_\psi}<0$ and $a_{2\Delta_\psi + 2j}>0$ for $j\in\mathbb{N}$ as required from an extremal functional. It is interesting to study the behaviour of $a_n$ for fixed $\Delta_\psi$ and $n\gg1$. Note first that
\bee
\alpha_{\Delta_\psi}(n,m) &= O\left(n^{-4m+2}\right)\\
\beta_{\Delta_\psi}(n,m) &= O\left(n^{-4m}\right)
\eee
as $n\rightarrow\infty$. However, due to delicate cancellations among all the terms in the sum in \eqref{eq:OmegaInteger}, $a_{n}$ decays as
\be
a_{n} = O\left(n^{-4\Delta_\psi-1}\right)\quad\textrm{as }n\rightarrow\infty\,.
\ee
In particular, this means that the sum \eqref{eq:hXInteger} converges to a smooth integral kernel $h(x)$ for $x\in(0,1)$ for any $\Delta_\psi\in\mathbb{N}$ and the convergence improves as $\Delta_\psi$ increases.

We can also find a closed formula for the action of $\omega_{\Delta_\psi}$ on $F_\Delta$ for any $\Delta\geq 0$. Orthogonality \eqref{eq:orthogonality} implies
\be
\omega_{\Delta_\psi}(\Delta) =
\begin{dcases}
\frac{\Gamma(2\Delta)}{\Gamma(\Delta)^2}\Omega_{\Delta_\psi}(\Delta)\quad&\textrm{for }\Delta\in\mathbb{N}\textrm{ even },\,\Delta\ge 2\Delta_\psi\\
0\quad&\textrm{for }\Delta\in\mathbb{N}\textrm{ odd },\,\Delta\ge 2\Delta_\psi+1\,.
\end{dcases}
\ee
The simplest meromorphic function of $\Delta$ with no other zeros or poles for $\Delta\geq 2\Delta_\psi$ is
\be
\omega_{\Delta_\psi}(\Delta) = \frac{\Gamma(2\Delta)}{\Gamma(\Delta)^2}\cos^2\left(\frac{\pi\Delta}{2}\right)\Omega_{\Delta_\psi}(\Delta)\,,
\label{eq:omegaIntAct}
\ee
which turns out to be the right answer.
%%%%%%%%%
\begin{figure}
\centering
\includegraphics[width=0.9\textwidth]{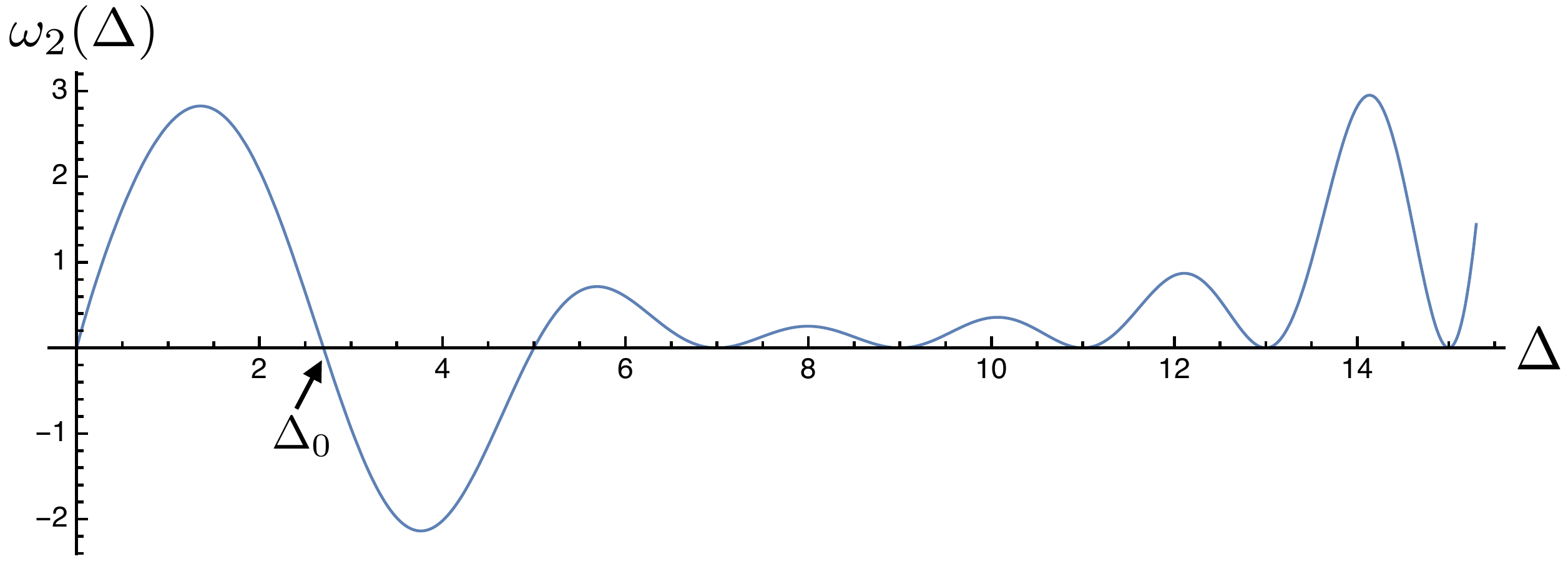}
\caption{The action of the extremal functional for $\Delta_\psi=2$, given by equation \eqref{eq:omegaIntAct}. The location $\Delta_0$ of the only zero in $0<\Delta<2\Delta_{\psi}+1$ will satisfy as $\Delta_0/\Delta_\psi\rightarrow\sqrt{2}$ as $\Delta_{\psi}\rightarrow\infty$.}\label{fig:omegaTwoAct}
\end{figure}
%%%%%%%%%
Figure \ref{fig:omegaTwoAct} shows the action of the extremal functional for $\Delta_\psi = 2$. $\Omega_{\Delta_\psi}(\Delta)$ is positive with no zeros or poles for $\Delta > 2\Delta_\psi + 1$. The only zeros in this region are thus the double zeros coming from $\cos^2(\pi\Delta/2)$. $\Omega_{\Delta_\psi}(\Delta)$ has a simple pole at $\Delta = 2\Delta_\psi+1$ coming from the $\beta_{\Delta_\psi}(\Delta,\Delta_\psi)$ summand in \eqref{eq:OmegaInteger}, leading to a simple zero at that location. All the double zeros of $\cos^2(\pi\Delta/2)$ in the region $0<\Delta<2\Delta_\psi$ are cancelled by double poles of $\Omega_{\Delta_\psi}(\Delta)$. The only zero of $\omega_{\Delta_\psi}(\Delta)$ in this region occurs at some non-integer value $\Delta_0$. We will see in section \ref{sec:ads} that
\be
\frac{\Delta_0}{\Delta_\psi} \rightarrow \sqrt{2}\quad\textrm{as }\Delta_\psi\rightarrow\infty\,,
\ee
which will be crucial to make contact with the flat-space limit. The functional $\omega_{\Delta_\psi}$ satisfies all the properties \eqref{eq:1DFunctionalProperties}, thus establishing rigorously that the 1D bootstrap bound on the gap is $2\Delta_\psi+1$ for any $\Delta_\psi\in\mathbb{N}$.

We were not able to find a closed form for the kernel defined by the sum \eqref{eq:hXInteger}. The kernel is observed to have the following behaviour near $x = 0$
\be
h(x) \sim h_1(x) + \log (x)h_2(x)\,,
\ee
where $h_1(x)$ and $h_2(x)$ are analytic at $x=0$ with the leading behaviour
\bee
h_1(x) &= a_0 + a_1 x +\ldots\\
h_2(x) &= b_{0}x^{2\Delta_\psi} + b_{1}x^{2\Delta_\psi+1} + \ldots\,.
\label{eq:h1h2}
\eee
Recall that for general $\Delta$, the action of $\omega$ must be defined via the contour integral \eqref{eq:functionalAction2}. The branch cut of $h(x)$ along $x\in(-\infty,0)$ arising from the infinite sum over $p_n(x)$ leads to a dependence on deformations of $\Gamma$. This dependence appears because the infinite sum over $n$ and the analytic continuation in $x$ do not commute. The correct choice reproducing the answer \eqref{eq:omegaIntAct} is one where the contour intersects the negative real axis arbitrarily close to $x=0$. We can not take the contour all the way to $x=0$ because of the $x^{\Delta-2\Delta_\psi}$ singularity in $F_\Delta(x)$. The $x^{2\Delta_\psi}$ supression of $h_2(x)$ seen in \eqref{eq:h1h2} guarantees that the value of the contour integral converges as the intercept approaches $x=0$.

%%%%%%%%%%%%%%%%%%
\subsection{Fixing the remaining redundancy}\label{ssec:fixred}
As explained in Section \ref{sec:commentsfunctional}, the extremal functional for any $\Delta_\psi$ can likely be expressed in the basis of derivatives with respect to the Zhukovsky variable $y$
\be
\omega_{\Delta_\psi} = \sum\limits_{j=1}^{\infty}\frac{b_j}{(2j-1)!}\left.\frac{d^{2j-1}}{dy^{2j-1}}\right|_{y=0}\,,
\label{eq:yDerFun}
\ee
where $b_j\in\mathbb{R}$ depend on $\Delta_\psi$. Requiring the existence of this representation will fix the redundancy in $h(x)$ described in subsection \ref{ssec:lindep}. Recall from section \ref{sec:commentsfunctional} that (ignoring the singularity at $\theta=0$ present only for small $\Delta$) \eqref{eq:yDerFun} can be expressed as the integral
\be
\omega_{\Delta_\psi}(\Delta) =\frac{4}{\pi}\int\limits_0^{\pi/2}\!\d\theta\,g_{\Delta_\psi}(\theta) \Im[F_\Delta\left(e^{i\theta}\right)]\,,
\ee
where
\be
g_{\Delta_\psi}(\theta) = \sum\limits_{j=1}^{\infty}b_{j}\sin\left[(2j-1)\theta\right]\,.
\ee
In other words, $b_{j}$ are simply the Fourier coefficients of the integral kernel constructed above. Coordinates $x$ and $\theta$ are related through
\be
x=\tan^2\left(\frac{\theta}{2}\right)\,,
\ee
and the Fourier coefficients can be obtained from the integral kernel $h(x)(1-x)^{-2\Delta_\psi}$ via
\be
b_j = \frac{1}{\pi}\int\limits_0^1\!\d x\sin\left[(2j-1)\theta(x)\right]h(x)(1-x)^{-2\Delta_\psi}\,.
\label{eq:bjFourier}
\ee
According to the results of the previous subsection, $h(x)$ remains nonzero as $x\rightarrow 1$. Therefore, the integral diverges at $x=1$ for any $\Delta_\psi\in\mathbb{N}$ and the integral kernel seems not to have a Fourier expansion. Fortunatelly, this problem can be amended by recalling there is an ambiguity in $h(x)$, described in subsection \ref{ssec:lindep}. The behaviour of $h(x)$ defined by \eqref{eq:hXInteger} and \eqref{eq:hInteger} near $x=1$ is
\be
h(x)\sim - h_1(1-x) - h_2(1-x)\log(1-x)\,,
\ee
with $h_{1,2}(x)$ as in \eqref{eq:h1h2}. Therefore, only $h_1(1-x)$ up to $O((x-1)^{2\Delta_\psi-1})$ contributes to the singularity of \eqref{eq:bjFourier}, while the logarithmic term does not contribute to the singularity since it is sufficiently supressed. We must now ask whether there exists a polynomial $c(x)\in\ker\varphi$ such that
\be
-h_1(1-x) + c(x) = O((x-1)^{2\Delta_\psi})
\ee
as $x\rightarrow 1$. This is a priori an overconstrained problem since we need to cancel $2\Delta_\psi$ independent coefficients of $h_1$ using a polynomial taken from the space $\ker\varphi$ of dimension $\Delta_\psi$. However, we found that it was possible for all $1\leq\Delta_\psi\leq5$ and therefore it is likely possible in general. We were not able to find a closed formula for $c(x)$ for general $\Delta_\psi\in\mathbb{N}$. Listed below are some low-lying examples of $c(x)$
\bee
\Delta_\psi=1:\quad&c(x) = -\frac{3}{8}\\
\Delta_\psi=2:\quad&c(x) = \frac{15}{16}x-\frac{2505}{1024}(x-1)^2\\
\Delta_\psi=3:\quad&c(x) = \frac{35}{8}x^2-\frac{6055}{1024}x(x-1)^2-\frac{418985}{65536}(x-1)^4\,.
\eee
Clearly $c(x)\in\ker\varphi$ with $\ker\varphi$ given by \eqref{eq:ker} in all these examples. The Fourier coefficients can now be derived as
\be
b_j = \frac{1}{\pi}\int\limits_0^1\!\d x\sin\left[(2j-1)\theta(x)\right]\left[h(x)+c(x)\right](1-x)^{-2\Delta_\psi}\,.
\ee
The extremal functionals coming from numerical bootstrap in the derivative basis were checked to tend to these analytic predictions for $\Delta_{\psi}=1$ as $N_{\textrm{max}}$ was increased although the convergence rate was slower compared to $\Delta_{\psi}=1/2$ presented in Figure \ref{fig:yDerCoefs0p5}.

%%%%%%%%%%%%%%%%%%%
\subsection{Extremal functionals for $\Delta_\psi\in\mathbb{N}-\frac{1}{2}$}\label{ssec:FunctionalsHalfInt}
Let us move on to describe the extremal functionals in the case $\Delta_\psi\in\mathbb{N}-1/2$. It follows from the result of subsection \ref{ssec:properties} that only $a_n$ with $n$ odd are nonvanishing for $n\geq 2\Delta_\psi$. Analogously to \eqref{eq:hXInteger}, we might hope that $h(x)$ can be expanded using only $p_n(x)$ with $n$ odd. However, the space of functions $s(\Delta,n)$ with $1\leq n\leq2\Delta_\psi - 1$ is spanned by the same functions with $n$ restricted to be even, but not $n$ restricted to be odd. Indeed, it turns out that in order for the functional to have double zeros at the right locations, i.e. for \eqref{eq:MatrixEq} to hold, some $a_n$ with $1\leq n\leq2\Delta_\psi - 1$ and $n$ even must be nonvanishing. We can write
\be
h(x) = \tilde{h}(x) + c(x)\,,
\ee
where
\be
\tilde{h}(x) = \sum\limits_{n\in2\mathbb{N}-1} \tilde{a}_{n}p_{n}(x)\,,
\label{eq:hXInteger2}
\ee
and $c(x)$ is a polynomial of degree at most $2\Delta_\psi - 2$. We expect the extremal functional can still be represented by a derivative series \eqref{eq:yDerFun}, meaning
\be
h(x)(1-x)^{-2\Delta_\psi}
\ee
has at most a logarithmic singularity at $x=1$. This requirement fixes $c(x)$ for any choice of the sequence $\tilde{a}_{2j-1}$. We will now present a formula for $\tilde{a}_{n}$ such that the corresponding $h(x)$ with $c(x)$ fixed by this requirement satisfies \eqref{eq:MatrixEq}. First, define
\bee
\tilde{\alpha}_{\Delta_\psi}(\Delta,m) &=
-\left[2 m \left(\Delta _{\psi }- m-1\right)+\Delta_\psi\right]
\frac{\pi\Gamma (4 m+1)\Gamma\left(m\right)^2
\Gamma \left(\frac{\Delta+1}{2}\right)^2
\Gamma\left(\frac{\Delta}{2}-m\right)^2}
{2^{8 m+1}\Gamma\left(m+\frac{1}{2}\right)
\Gamma\left(m+\frac{3}{2}\right)
\Gamma \left(\frac{\Delta}{2}\right)^2
\Gamma \left(\frac{\Delta+1}{2}+m\right)^2}\\
\tilde{\beta}_{\Delta_\psi}(\Delta,m) &=
\left(\Delta_{\psi}-m-1\right)
\frac{\pi\Gamma (4 m+2)\Gamma\left(m+1\right)^2
\Gamma \left(\frac{\Delta+1}{2}\right)^2
\Gamma\left(\frac{\Delta}{2}-m\right)\Gamma\left(\frac{\Delta}{2}-m-1\right)}
{2^{8 m+2}
\Gamma\left(m+\frac{1}{2}\right)
\Gamma\left(m+\frac{3}{2}\right)
\Gamma \left(\frac{\Delta}{2}\right)^2
\Gamma \left(\frac{\Delta+1}{2}+m\right)\Gamma \left(\frac{\Delta+3}{2}+m\right)}\,.
\label{eq:vwtilde}
\eee
Use these to define $\Omega_{\Delta_\psi}(\Delta)$ for $\Delta_\psi\in\mathbb{N}-1/2$
\be
\Omega_{\Delta_\psi}(\Delta)=
-\left[\Delta(\Delta-1)+\Delta_\psi \right]\Psi'\left(\frac{\Delta}{2}\right)-2
-\sum\limits_{m=1}^{\Delta_\psi-\frac{1}{2}}\tilde{\alpha}_{\Delta_\psi}(\Delta,m)
-\sum\limits_{m=0}^{\Delta_\psi-\frac{1}{2}}\tilde{\beta}_{\Delta_\psi}(\Delta,m)
\ee
with $\Psi(z)$ defined in \eqref{eq:PsiFun}.

The formula for $\tilde{a}_{n}$ is
\be
\tilde{a}_n =
\begin{dcases}
(2n-1)\Omega_{\Delta_\psi}(n)\quad&\textrm{for }n\textrm{ odd}\\
0\quad&\textrm{for }n\textrm{ even}
\end{dcases}
\label{eq:hHalfInteger}
\ee
We were not able to find a closed formula for $c(x)$ completing $\tilde{h}(x)$ to the full integral kernel for any $\Delta_\psi$, but checked that $c(x)$ consistent with the constraints existed for all $1/2\leq\Delta_\psi\leq9/2$. Several low-lying examples follow
\bee
\Delta_\psi=\frac{1}{2}:\quad&c(x) = 0\\
\Delta_\psi=\frac{3}{2}:\quad&c(x) = \frac{35}{12}(x-2)\\
\Delta_\psi=\frac{5}{2}:\quad&c(x) = (x-2)\left[\frac{1001}{120}(x^2-x+1)+\frac{\pi^2}{10}(2x^2+x-1)\right]\,.
\label{eq:cpolys}
\eee
Analogously to the case with $\Delta_\psi\in\mathbb{N}$, the action of the extremal functionals on $F_\Delta$ with any $\Delta>0$ reads
\be
\omega_{\Delta_\psi}(\Delta) = \frac{\Gamma(2\Delta)}{\Gamma(\Delta)^2}\sin^2\left(\frac{\pi\Delta}{2}\right)\Omega_{\Delta_\psi}(\Delta)\,.
\label{eq:omegaHalfIntAct}
\ee
Discussion following \eqref{eq:omegaIntAct} concerning zeros $\omega_{\Delta_\psi}(\Delta)$ applies in this case too with obvious modifications. \textbf{In summary, we have constructed functionals which prove that the optimal bootstrap bound is $2\Delta_\psi + 1$ for $\Delta_\psi\in\mathbb{N}/2$.}

The resummation of \eqref{eq:hXInteger2} is simpler than in the case of integer $\Delta_{\psi}$. It appears $h(x)$ for $\Delta_\psi\in\mathbb{N}-1/2$ can always be written in terms of rational, $\log$ and $\mathrm{Li}_2$ functions. We present some closed formulas for $\tilde{h}(x)$ in Appendix \ref{app:closed}.

%%%%%%%%%%%%%%%%%%%%
\section{Emergence of $\textrm{AdS}$ physics at large $\Delta_\psi$}\label{sec:ads}
%%%%%%%%%%%%%%%%%%%%%%%
\subsection{A review of massive scattering in large $\textrm{AdS}_2$}
\begin{figure}
\centering
\includegraphics[width=0.75\textwidth]{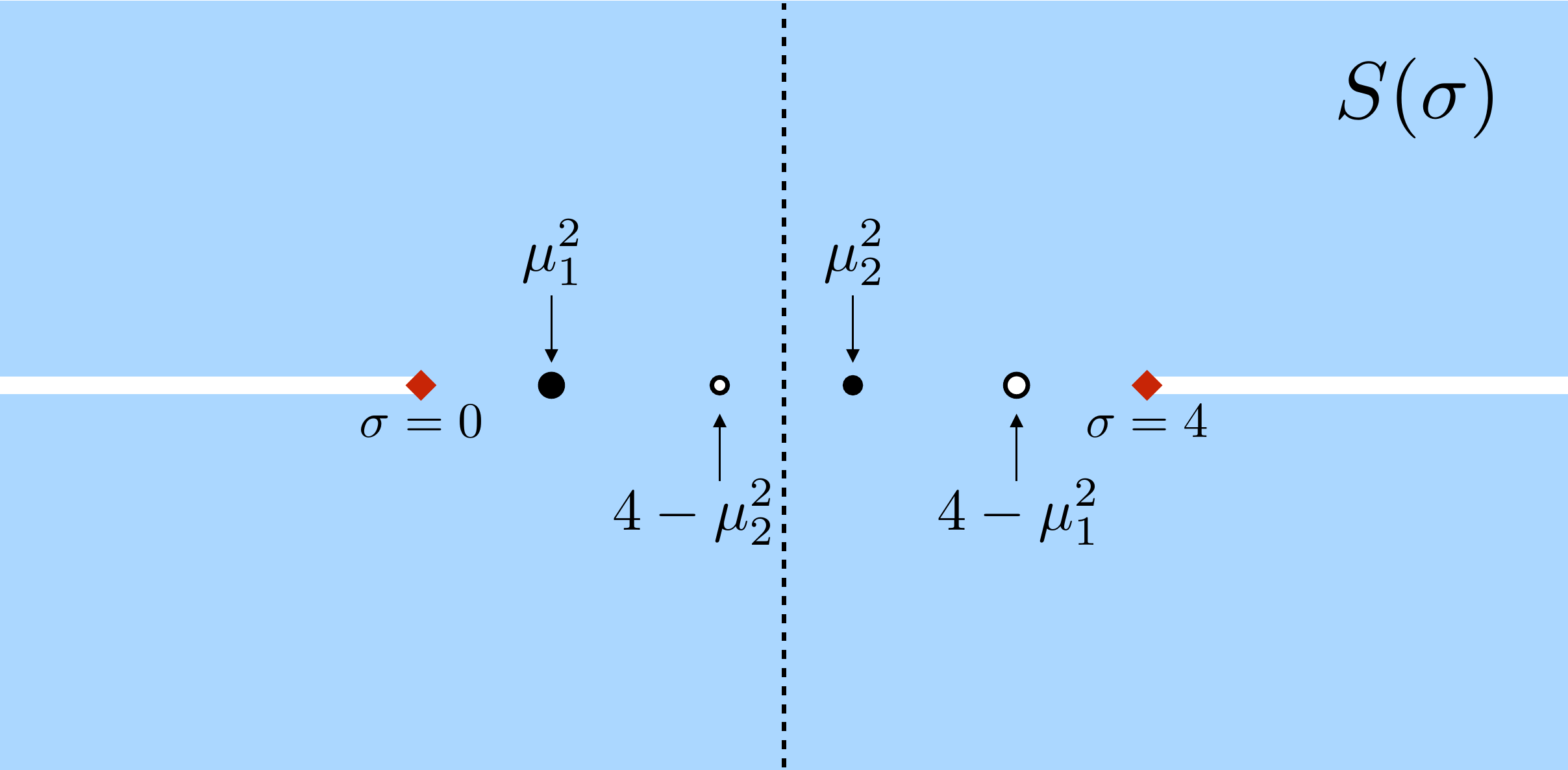}
\caption{The analytic structure of the S-matrix on the first sheet. There is a branch cut for real $\sigma>4$ corresponding to two-particle final states. Crossing symmetry implies $S(4-\sigma) = S(\sigma)$. The region $0<\sigma<4$ contains poles coming from bound states. Full and empty dots denote s- and t-channel poles respectively.}
\label{fig:SMatrix}
\end{figure}

It turns out that the extremal functional constructed in the previous section has a clear physical meaning for large $\Delta_\psi$ in terms scattering of massive particles in large $\textrm{AdS}_2$. As a first hint, we can notice that the location of the only zero $\Delta_0$ of $\omega_{\Delta_{\psi}}(\Delta)$ in the region $0<\Delta<2\Delta_\psi$ tends to
\be
\lim_{\Delta_\psi\rightarrow\infty}\frac{\Delta_0}{\Delta_\psi}\rightarrow\sqrt{2}\,,
\ee
which corresponds to the point fixed by the crossing symmetry of the flat-space S-matrix. We begin by reviewing aspects of two-dimensional scattering and its holographic dictionary. More details and derivations can be found in \cite{SMatrix1}. Consider a $2\rightarrow 2$ scattering amplitude of identical particles of mass $m_\psi$ in (1+1)D flat spacetime. The amplitude is fully described by the S-matrix $S(\sigma)$, where
\be
\sigma = \frac{(p_1+p_2)^2}{m_\psi^2}
\ee
is a dimensionless version of the usual Mandelstam variable $s$. The analytic structure of the S-matrix is illustrated in Figure \ref{fig:SMatrix}. Physical scattering regime corresponds to $\sigma\geq 4$, and $S(\sigma)$ has a branch cut there. The branch cut is of the square root type. In the extreme non-relativistic regime $\sigma \rightarrow 4$ the particles become free, and thus the leading behaviour is
\be
S(\sigma)=\pm 1 + \alpha\sqrt{4-\sigma} + O(4-\sigma)\,,
\label{eq:sExp}
\ee
where the upper, lower sign corresponds to bosons, fermions respectively and $\alpha\in\mathbb{R}$. Unitarity implies
\be
|S(\sigma)| \leq 1\quad\textrm{for }\sigma\geq 4\,,\sigma\in\mathbb{R}\,.
\ee
%%%%%%%%%
%%%%%%%%%
The S-matrix also satisfies crossing symmetry
\be
S(4-\sigma) = S(\sigma)\,.
\ee
Let us assume the scattered particle is the lightest particle of the theory. In that case, the only other singularities of $S(\sigma)$ on the first sheet are simple poles on the real axis in $0<\sigma<4$, coming from bound states and located at $\sigma_j = \mu_j^2$, as well as the corresponding $t$-channel poles located at $\sigma_j = 4-\mu_j^2$, where
\be
\mu_j = \frac{m_j}{m_\psi}
\ee
and $m_j$ is the bound state mass. The two kinds of poles can be distinguished by the sign of their residue
\be
S(\sigma)\sim
\begin{dcases}
 -\mathcal{J}_j\frac{g_{j}^2}{\sigma - \mu_{j}^2}\quad&\textrm{near }\sigma=\mu_j^2\\
\mathcal{J}_j\frac{g_{j}^2}{\sigma - (4-\mu_{j}^2)}\quad&\textrm{near }\sigma=4-\mu_j^2\,,
\end{dcases}
\label{eq:SMatRes}
\ee
where $g_j\in\mathbb{R}$ is the effective three-point coupling between the external particles and the bound state, and $\mathcal{J}_j$ is the positive prefactor
\be
\mathcal{J}_j = \frac{1}{2\mu_j\sqrt{4-\mu_j^2}}\,.
\ee

Placing the theory in $\textrm{AdS}_2$ of radius $R$ defines a family of 1D CFTs parametrized by $R$. Bulk masses and boundary scaling dimensions of primary operators are related by
\be
(m_{\mathcal{O}} R)^2 = \Delta_{\mathcal{O}}(\Delta_{\mathcal{O}} - 1)\,,
\ee
When we send $R\rightarrow\infty$, all scaling dimensions of a theory whose bulk dual is a massive QFT tend to infinity. Their ratios tend to the ratios of the corresponding masses
\be
\lim_{\Delta_\psi\rightarrow\infty} \frac{\Delta_j}{\Delta_\psi} = \mu_j\,.
\label{eq:DelRat}
\ee
The $2\rightarrow2$ scattering corresponds to a four-point function of primary operators $\psi(x)$ sourcing the external particle. Primary operators appearing in the $\psi\times\psi$ OPE correspond to intermediate states of the scattering process. Those with $\Delta_{\mathcal{O}}\lesssim2\Delta_\psi$ play the role of bound states, while those with $\Delta_{\mathcal{O}}\gtrsim2\Delta_\psi$ correspond to two-particle states. The flat-space physics governs the leading behaviour of the CFT data as $\Delta_\psi\rightarrow\infty$. The flat-space scattering amplitude can be recovered as a specific limit of the boundary Mellin amplitude \cite{MellinJoao,SMatrix1}. For example, the leading behaviour of the OPE coefficient $c_{\psi\psi\mathcal{O}_j}$ corresponding to a bound state of mass $m_j=\mu_j m_\psi$ takes the form
\be
(c_{\psi\psi\mathcal{O}_j})^2\sim
\frac{2 \sqrt{\pi } g_j^2}
{\mu _j^{3/2}(4-\mu_j ^2)\left(\mu _j+2\right)}
\sqrt{\Delta_\psi}\left[v(\mu _j)\right]^{-\Delta_\psi}\quad
\textrm{as }
\Delta_\psi\rightarrow\infty
\,,
\label{eq:cBound}
\ee
where
\be
v(\mu) = 
\frac{4^{2+\mu}}{|\mu-2|^{2-\mu } (\mu+2)^{2+\mu}}\,,
\label{eq:vmu}
\ee
and $g_j$ is defined in \eqref{eq:SMatRes}. We inserted the absolute value around $\mu-2$ for future convenience. In the bound state region $0<\mu<2$, we have $v(\mu)>1$, and thus $c_{\psi\psi\mathcal{O}_j}$ is exponentially supressed in the large $\Delta_\psi$ limit. This supression is coming from the amplitude for the massive particles to propagate across an increasingly large distance in $\textrm{AdS}$, as explained in \cite{Raul1,Raul2}. We will be able to recover the exponential supression including the precise dependence of the exponent on $\mu$ from conformal bootstrap.

Consider now the primary operators in the $\psi\times\psi$ OPE coming from the two-particle states. When the bulk theory is that of free real bosons or fermions, there is an exact formula for the OPE coefficients
\be
(c_{\psi\psi\mathcal{O}}^{\textrm{free}})^2 =
\frac{2 \Gamma (\Delta_{\mathcal{O}})^2 \Gamma\left(\Delta_{\mathcal{O}} +2 \Delta _{\psi}-1\right)}{\Gamma (2 \Delta_{\mathcal{O}} -1)\Gamma \left(2 \Delta _{\psi }\right)^2\Gamma \left(\Delta_{\mathcal{O}} -2 \Delta _{\psi }+1\right)}\,,
\ee
and the scaling dimensions are $\Delta_{\mathcal{O}} = 2\Delta_\psi + n$, where $n$ is an even, odd non-negative integer for bosons, fermions respectively. For a general theory in $\textrm{AdS}_2$, define the spectral density
\be
\rho_{\Delta_\psi}(\mu) = \sum_{\mathcal{O}\in\psi\times\psi}(c_{\psi\psi\mathcal{O}})^2\delta\left(\mu-\frac{\Delta_{\mathcal{O}}}{\Delta_\psi}\right)\,.
\label{eq:SpecRho}
\ee
It can be shown \cite{SMatrix1} that $\rho_{\Delta_\psi}(\mu)$ is universal in the flat-space limit in the sense that it tends to the asymptotic spectral density of free fields, namely
\be
\rho_{\Delta_\psi}(\mu)\sim
\tilde{\rho}_{\Delta_\psi}(\mu)=\frac{4 \sqrt{\mu }}{\sqrt{\pi } \sqrt{\mu -2} \left(\mu+2\right)^{\frac{3}{2}}}
\sqrt{\Delta_\psi}
\left[v(\mu)\right]^{-\Delta_\psi}
\quad
\textrm{as }
\Delta_\psi\rightarrow\infty\,,
\label{eq:SpecDens}
\ee
with $v(\mu)$ again given by \eqref{eq:vmu}. Equation \eqref{eq:SpecDens} is valid in the sense of distributions when acting on smooth functions of $\mu$.

Finally, it is also possible to recover the S-matrix from the shifts of scaling dimension of two-particle states compared to their free-field positions. To this end, define the following smeared average of an arbitrary function $f(\mu,\Delta_\psi)$
\be
\langle f(\mu,\Delta_\psi)\rangle_{\epsilon} =
\frac{
\int\limits_{\mu-\epsilon}^{\mu+\epsilon}
\d\nu\rho_{\Delta_{\psi}}(\nu)\left[v(\nu)\right]^{\Delta_\psi}f(\nu,\Delta_\psi)
}
{
\int\limits_{\mu-\epsilon}^{\mu+\epsilon}
\d\nu\rho_{\Delta_{\psi}}(\nu)\left[v(\nu)\right]^{\Delta_\psi}\,,
}
\ee
where $\rho_{\Delta_{\psi}}(\nu)$ is the exact spectral density at finite $\Delta_\psi$. The factor $\left[v(\nu)\right]^{\Delta_\psi}$ cancels the fast variation of $\rho_{\Delta_\psi}(\nu)$ with $\nu$ when $\Delta_\psi\rightarrow\infty$. The S-matrix for $\sigma\geq4$ can now be recovered through the formula
\be
S(\mu^2) = \lim_{\epsilon\rightarrow 0}\lim_{\Delta_\psi\rightarrow\infty}\left\langle
e^{-i\pi(\mu-2)\Delta_{\psi}}
\right\rangle_{\epsilon}\,,
\label{eq:SMatFromRho}
\ee
where the order of limits is important. In other words, $S(\mu^2)$ is simply the large $\Delta_\psi$ limit of the average value of $e^{-i\pi(\Delta_{\mathcal{O}}-2\Delta_\psi)}$ over all primaries with $\Delta_{\mathcal{O}}\sim\mu\Delta_\psi$, weighted by $(c_{\psi\psi\mathcal{O}}/c_{\psi\psi\mathcal{O}}^{\textrm{free}})^2$.

%%%%%%%%%%%%%%%%%%%%%%%
\subsection{$\textrm{AdS}_2$ physics from crossing in a 1D CFT}
We will now show how some of the features pertaining to the scattering of massive particles in large $\textrm{AdS}_2$ presented in the previous subsection follow from crossing in the 1D CFT living at the boundary. We will also derive a simple sum rule for OPE coefficients of primary operators corresponding to two-particle states produced at rest.

\begin{figure}
\centering
\includegraphics[width=\textwidth]{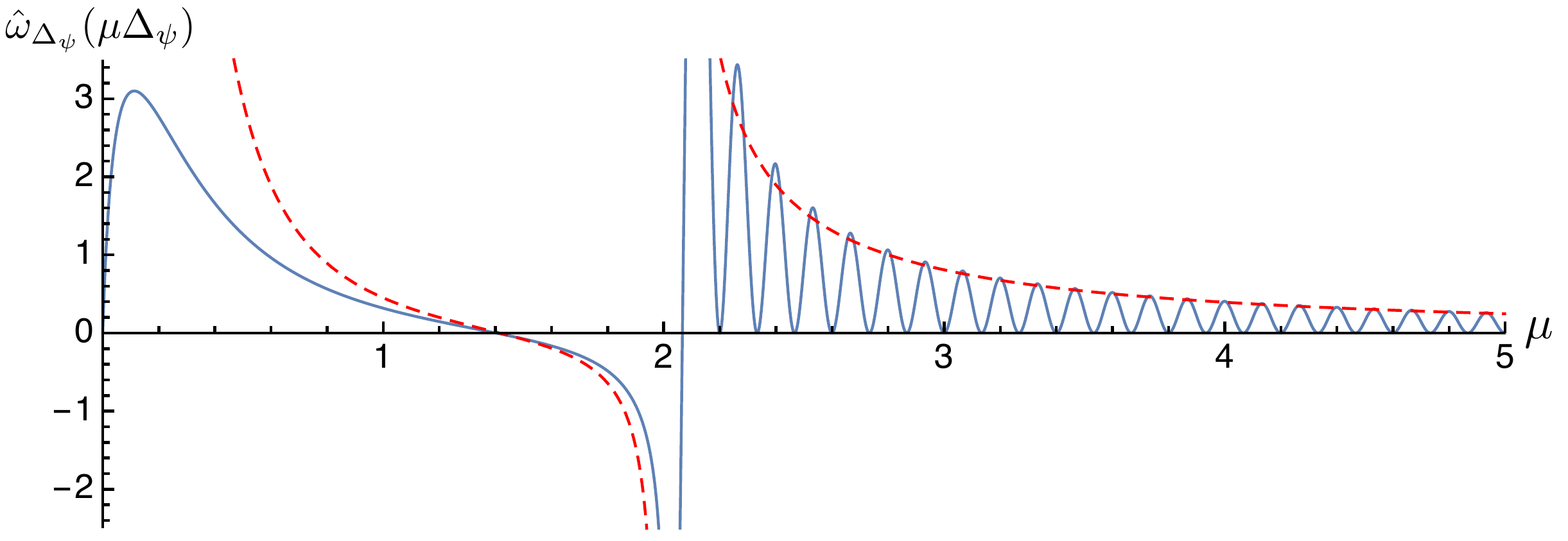}
\caption{The analytic extremal functional for $\Delta_\psi=15$. The blue curve represents $\hat{\omega}_{\Delta_\psi}(\mu \Delta_\psi) = \left[v(\mu)\right]^{-\Delta_\psi}\omega_{\Delta_\psi}(\mu \Delta_\psi)$ with $v(\mu)$ given by \eqref{eq:vmu} and $\omega_{\Delta_\psi}(\Delta)$ given by \eqref{eq:omegaIntAct}. $\hat{\omega}_{\Delta_\psi}(\mu \Delta_\psi)$ in the region $0<\mu<2$ converges to the red dashed curve given by $\sqrt{2}\pi^{-1}\mu^{-5/2}(\mu-2)^{-1}(\mu^2-2)$ as $\Delta_\psi\rightarrow\infty$. The functional is oscillatory in the region $\mu>2$ with evenly spaced double zeros that condense as $\Delta_\psi\rightarrow\infty$. The height of peaks converges to the red dashed curve given by $4\sqrt{2}\pi^{-1}\mu^{-5/2}(\mu-2)^{-1}(\mu^2-2)$.}
\label{fig:omegaHatD15}
\end{figure}

Consider a unitary solution to the bootstrap equation in 1D
\be
\sum\limits_{\mathcal{O}\in\psi\times\psi}(c_{\psi\psi\mathcal{O}})^2F_{\Delta_\mathcal{O}}(z) = 0\,,
\ee
with $(c_{\psi\psi\mathcal{O}})^2>0$. We can apply the extremal functional constructed in section \ref{sec:higherdelta} to get a single equation
\be
\sum\limits_{\mathcal{O}\in\psi\times\psi}(c_{\psi\psi\mathcal{O}})^2\omega_{\Delta_\psi}(\Delta_{\mathcal{O}}) = 0\,.
\label{eq:crossingOmega}
\ee
When the solution to crossing corresponds to the free massive real fermion in $\textrm{AdS}_2$, the last equation is automatically satisfied since $\omega_{\Delta_\psi}(\Delta_{\mathcal{O}})=0$ for any $\mathcal{O}\in\psi\times\psi$. However, in general it represents a universal constraint valid on any solution of crossing. In order to make contact with massive QFT in $\textrm{AdS}_2$, let us assume we have a family of solutions where all dimensions scale linearly with $\Delta_\psi$, i.e. that $\Delta_{\mathcal{O}}\sim \mu_{\mathcal{O}}\Delta_\psi$ with $\mu_{\mathcal{O}}$ fixed as $\Delta_\psi\rightarrow\infty$. We would like to understand the leading behaviour of \eqref{eq:crossingOmega} as $\Delta_\psi\rightarrow\infty$. The functional $\omega_{\Delta_\psi}(\mu\Delta_\psi)$ with large $\Delta_\psi$ exhibits very different behaviour for $0<\mu<2$ and $\mu>2$, as illustrated in Figure \ref{fig:omegaHatD15}.
%%%%%%%%%
%%%%%%%%%

\begin{figure}
\centering
\includegraphics[width=0.75\textwidth]{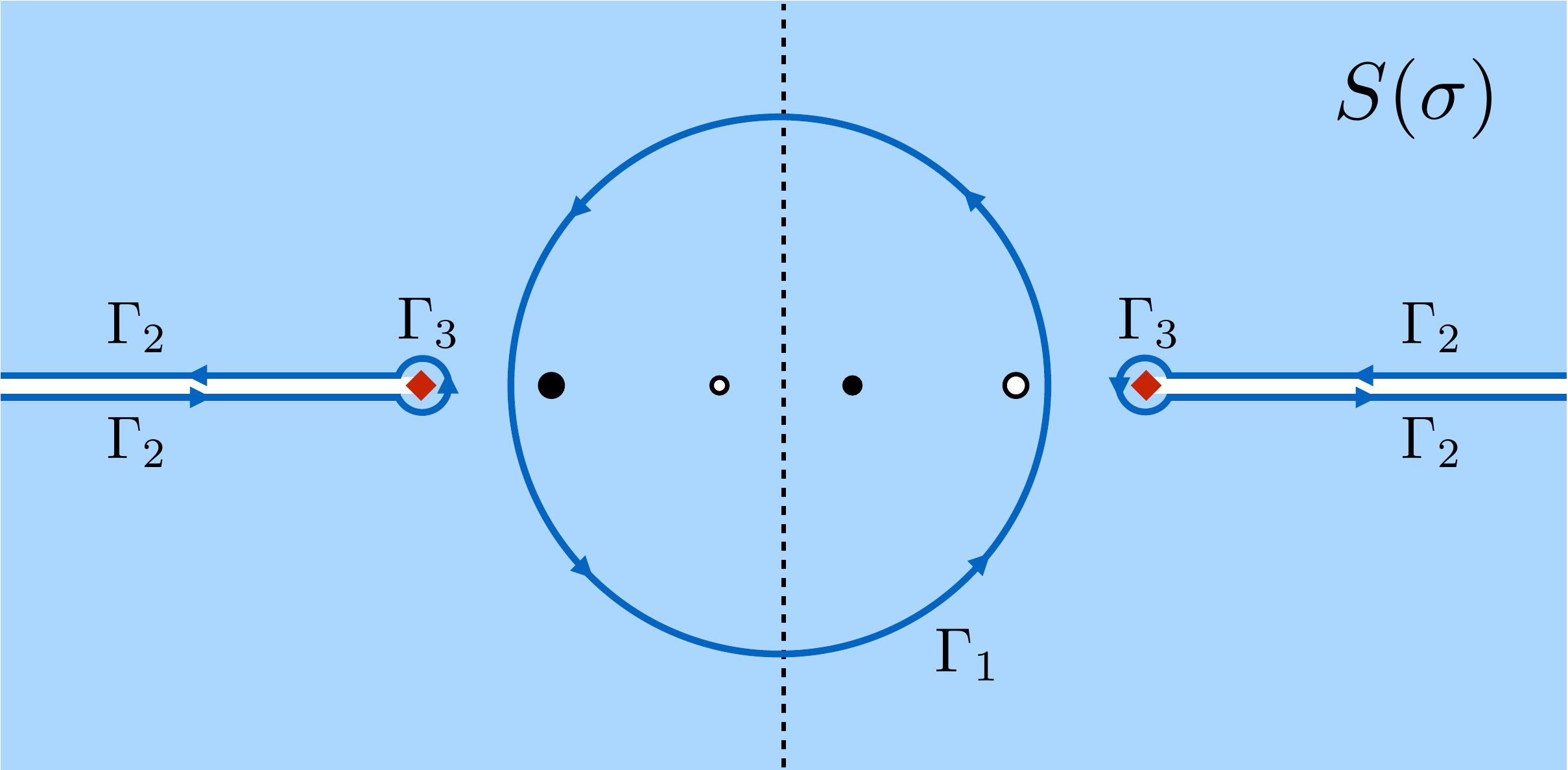}
\caption{Contour integrals describing the contributions of various states to the crossing equation \eqref{eq:crossingOmega} at the leading order as $\Delta_\psi\rightarrow\infty$. $\Gamma_1$, $\Gamma_2$ and $\Gamma_3$ give the contribution of bound states, two-particle states and $\mu=2$ states respectively. The integrand takes the form $(\sigma-2)\sigma^{-3/2}(4-\sigma)^{-3/2}[S(\sigma)+1]$. Conformal bootstrap at the leading order in $1/\Delta_\psi$ is equivalent to the total contour integral being zero, in other words to the analyticity of the S-matrix away from the real axis.}
\label{fig:ContourS}
\end{figure}

Let us focus first on the region $0<\mu<2$. It is possible to show directly from \eqref{eq:omegaIntAct} that
\be
\omega_{\Delta_\psi}(\mu\Delta_\psi)\sim
\frac{\sqrt{2}\left(\mu^2-2\right)}{\pi \mu^{\frac{5}{2}}(\mu-2)}\left[v(\mu)\right]^{\Delta_\psi}
\quad\textrm{for }0<\mu<2
\label{eq:omegaBound}
\ee
as $\Delta_\psi\rightarrow\infty$ where $v(\mu)$ was defined in \eqref{eq:vmu}. Crucially, the functional grows exponentially with $\Delta_\psi$ with the exponent governed by $v(\mu)$. It follows that any two operators $\mathcal{O}_j$, $\mathcal{O}_k$ with $0<\mu_{j,k}<2$ that both contribute to \eqref{eq:crossingOmega} at the leading order as $\Delta_\psi\rightarrow\infty$ must have OPE coefficients related by
\be
\frac{(c_{\psi\psi\mathcal{O}_j})^2}{(c_{\psi\psi\mathcal{O}_k})^2} \sim
\left[\frac{v(\mu_j)}{v(\mu_k)}\right]^{-\Delta_\psi}
\ee
up to a prefactor independent of $\Delta_\psi$. This is consistent with the exponential supression of $c_{\psi\psi\mathcal{O}}$ when $\mathcal{O}$ corresponds to a bound state of two $\psi$ particles in the flat space limit, seen in \eqref{eq:cBound}. \textbf{The extremal functional thus provides a universal CFT justification of the exponential decay of bound state OPE coefficients}. Assuming the full expression \eqref{eq:cBound}, we can evaluate the contribution of a single bound state to \eqref{eq:crossingOmega}
\be
(c_{\psi\psi\mathcal{O}_j})^2\omega_{\Delta_\psi}(\Delta_{\mathcal{O}_j}) =
2\sqrt{\frac{8\Delta_\psi}{\pi}}\mathrm{Res}_{\sigma=\mu_j^2}\left[\frac{(\sigma-2)}{\sigma^{\frac{3}{2}}(4-\sigma)^{\frac{3}{2}}}S(\sigma)\right]+O(1)\,,
\ee
i.e. the bound states contribute at $O(\sqrt{\Delta_\psi})$. Since the function in square brackets is odd under $\sigma\leftrightarrow 4-\sigma$, and remembering that every s-channel pole has its t-channel counterpart, we can rewrite the contribution of all bound states to \eqref{eq:crossingOmega} at $O(\sqrt{\Delta_\psi})$ as a contour integral in the $\sigma$ plane along a contour $\Gamma_1$ surrounding all the poles on the real axis, as illustrated in Figure \ref{fig:ContourS}
\be
\sum\limits_{\mathcal{O}:\,\mu_{\mathcal{O}}<2}(c_{\psi\psi\mathcal{O}})^2\omega_{\Delta_\psi}(\Delta_{\mathcal{O}})=
\frac{1}{2\pi i}\sqrt{\frac{8\Delta_\psi}{\pi}}\oint\limits_{\Gamma_1}
\frac{(\sigma-2)}{\sigma^{\frac{3}{2}}(4-\sigma)^{\frac{3}{2}}}\left[S(\sigma)+1\right]
\d \sigma + O(1)\,,
\label{eq:BSContr}
\ee
where we added 1 to the S-matrix for future convenience without affecting the result. We will assume that the scattered particles are fermions, so that $S(0)=S(4)=-1$. The bosonic cases $S(0)=S(4)=1$ can presumably be treated analogously. 

Let us now study the asymptotic behaviour of the functional for $\mu>2$. As illustrated in Figure \ref{fig:omegaHatD15}, it is oscillatory with frequency proportional to $\Delta_\psi$. The asymptotics can be found in a closed form
\be
\omega_{\Delta_\psi}(\mu\Delta_\psi)\sim
4\cos^2\left(\frac{\pi\mu\Delta_\psi}{2}\right)\frac{\sqrt{2}\left(\mu^2-2\right)}{\pi \mu^{\frac{5}{2}}(\mu-2)}\left[v(\mu)\right]^{\Delta_\psi}
\quad\textrm{for }\mu>2\,,
\label{eq:omegaFlatTwoP}
\ee
where we can see the same exponential behaviour once again. Let us describe the spectrum of our solution to crossing for $\mu>2$ using the spectral density $\rho_{\Delta_\psi}(\mu)$, defined in \eqref{eq:SpecRho}. The contribution of the $\mu>2$ operators to \eqref{eq:crossingOmega} can be written as an integral
\be
\sum\limits_{\mathcal{O}:\,\mu_{\mathcal{O}}>2}(c_{\psi\psi\mathcal{O}})^2\omega_{\Delta_\psi}(\Delta_{\mathcal{O}})
=
\int\limits_{2}^{\infty}\!\rho_{\Delta_\psi}(\mu)\omega_{\Delta_\psi}(\mu\Delta_\psi)\d\mu\,.
\label{eq:CrossTwoPcle}
\ee
Assuming that states with any $\mu$ contribute to \eqref{eq:crossingOmega} at the leading order as $\Delta_\psi\rightarrow\infty$, we arrive at the same exponential dependence of $\rho_{\Delta_\psi}$ on $\Delta_\psi$ as the one corresponding to two-particle states in $\textrm{AdS}_2$, see formula \eqref{eq:SpecDens}. Let us now evaluate \eqref{eq:CrossTwoPcle} using the asymptotics \eqref{eq:omegaFlatTwoP} and assuming the formula \eqref{eq:SpecDens}. Note that the oscillating prefactor in \eqref{eq:omegaFlatTwoP} can be rewritten using
\be
2\cos^2\left(\frac{\pi\mu\Delta_\psi}{2}\right) = \Re\left[e^{-i\pi(\mu-2)\Delta_\psi} + 1\right]\,,
\ee
where we used $\Delta_\psi\in\mathbb{N}$. The oscillating prefactor is clearly related to the S-matrix on the branch cut as computed by \eqref{eq:SMatFromRho}. Indeed, it is not too hard to show from \eqref{eq:SMatFromRho} that
\be
\sum\limits_{\mathcal{O}:\,\mu_{\mathcal{O}}>2}(c_{\psi\psi\mathcal{O}})^2\omega_{\Delta_\psi}(\Delta_{\mathcal{O}})
=
\frac{1}{\pi}\sqrt{\frac{8\Delta_\psi}{\pi}}\int\limits_{4}^{\infty}
\frac{(\sigma-2)}{\sigma^{\frac{3}{2}}(\sigma-4)^{\frac{3}{2}}}2\Re\left[S(\sigma)+1\right]\d\sigma+O(1)\,.
\ee
It is now useful to notice that for real $\sigma>4$
\be
2\frac{\Re\left[S(\sigma)+1\right]}{(\sigma-4)^{\frac{3}{2}}} = i \left[\frac{S(\sigma)+1}{(4-\sigma)^{\frac{3}{2}}}\right]_{\sigma-i\epsilon}^{\sigma+i\epsilon}\,,
\ee
and therefore the last integral can be written as the contour integral
\be
\sum\limits_{\mathcal{O}:\,\mu_{\mathcal{O}}>2}(c_{\psi\psi\mathcal{O}})^2\omega_{\Delta_\psi}(\Delta_{\mathcal{O}})=
\frac{1}{2\pi i}\sqrt{\frac{8\Delta_\psi}{\pi}}\oint\limits_{\Gamma_2}
\frac{(\sigma-2)}{\sigma^{\frac{3}{2}}(4-\sigma)^{\frac{3}{2}}}\left[S(\sigma)+1\right]
\d \sigma + O(1)\,,
\label{eq:2PcleContr}
\ee
where $\Gamma_2$ consists of four half-lines lying on the branch cuts, as depicted in Figure \ref{fig:ContourS}, and we used $S(4-\sigma)=S(\sigma)$ to duplicate the contour from $\sigma>4$ to $\sigma<0$.
%%%%%%%%%
%%%%%%%%%

We arrived at a contour integral of exactly the same function as in the case of bound states \eqref{eq:BSContr}, only the contour is different now. The integrand decays as $\sigma^{-2}$ so it would be tempting to deduce the validity of \eqref{eq:crossingOmega} at the leading order in $\Delta_\psi$ from analyticity of $S(\sigma)$ away from the real axis by deforming $\Gamma_1\cup\Gamma_2$ to the empty contour. However, this is not a legal operation because the integrand has poles at $\sigma=0,4$ and thus the contour integral picks up a non-zero contribution from the infinitesimal contour $\Gamma_3$ depicted in Figure \ref{fig:ContourS}. On the CFT side, this contribution is coming from operators with $\mu=2$. The conformal bootstrap equation will thus be satisfied at the leading order if and only if
\be
\sum\limits_{\mathcal{O}:\,\mu_{\mathcal{O}}=2}(c_{\psi\psi\mathcal{O}})^2\omega_{\Delta_\psi}(\Delta_{\mathcal{O}})
=
\frac{1}{2\pi i}\sqrt{\frac{8\Delta_\psi}{\pi}}\oint\limits_{\Gamma_3}
\frac{(\sigma-2)}{\sigma^{\frac{3}{2}}(4-\sigma)^{\frac{3}{2}}}\left[S(\sigma)+1\right]
\d s+O(1)\,.
\label{eq:SlowContrEq}
\ee
Our asymptotic formulas \eqref{eq:omegaBound}, \eqref{eq:omegaFlatTwoP} for $\omega_{\Delta_\psi}(\mu\Delta_\psi)$ break down when $\mu=2$ and need to be modified. Primary operators $\mathcal{O}$ with $\mu=2$ are precisely those for which $\Delta_{\mathcal{O}}-2\Delta_\psi$ remains finite as $\Delta_\psi\rightarrow\infty$. Denote
\be
\delta_{\mathcal{O}} =\lim_{\Delta_\psi\rightarrow\infty}(\Delta_{\mathcal{O}} - 2\Delta_\psi)\,.
\ee
The asymptotics of $\omega_{\Delta_\psi}$ is modified to become a power-law
\be
\omega_{\Delta_\psi}(2\Delta_\psi + \delta_{\mathcal{O}})\sim
-\frac{1}{\Gamma\left(\frac{1-\delta}{2}\right)\Gamma\left(\frac{3-\delta}{2}\right)}
\left(\frac{\Delta_\psi}{2}\right)^{-\delta_{\mathcal{O}}+1}
\ee
as $\Delta_\psi\rightarrow\infty$. The contour integral on the right-hand side of \eqref{eq:SlowContrEq} can be evaluated using expansion \eqref{eq:sExp}
\be
\frac{1}{2\pi i}\sqrt{\frac{8\Delta_\psi}{\pi}}\oint\limits_{\Gamma_3}
\frac{(\sigma-2)}{\sigma^{\frac{3}{2}}(4-\sigma)^{\frac{3}{2}}}\left[S(\sigma)+1\right]
\d s = -\sqrt{\frac{2\Delta_\psi}{\pi}}\alpha\,,
\ee
where $\alpha$ is the coefficient of the square-root term in \eqref{eq:sExp}. Equation \eqref{eq:SlowContrEq} now implies the following sum rule for the OPE coefficients of the $\mu=2$ states
\be
\sum\limits_{\mathcal{O}:\,\mu_{\mathcal{O}}=2}
\frac{1}{\Gamma\left(\frac{1-\delta_{\mathcal{O}}}{2}\right)\Gamma\left(\frac{3-\delta_{\mathcal{O}}}{2}\right)}
\left(\frac{\Delta_\psi}{2}\right)^{-\delta_{\mathcal{O}}+\frac{1}{2}}
(c_{\psi\psi\mathcal{O}})^2
=
\frac{2\alpha}{\sqrt{\pi}} + O\left(\Delta_\psi^{-1/2}\right)\,,
\label{eq:SlowSum}
\ee
In particular, the OPE coefficients should scale as
\be
(c_{\psi\psi\mathcal{O}})^2 \sim a_{\mathcal{O}}\Delta_{\psi}^{\delta_{\mathcal{O}}-1/2}
\label{eq:opescaling}
\ee
as $\Delta_{\psi}\rightarrow\infty$ for these operators. We have shown that provided \eqref{eq:SlowSum} holds, \textbf{the validity of the conformal bootstrap equation \eqref{eq:crossingOmega} at the leading order at large $\Delta_\psi$ follows from analyticity of the flat-space S-matrix away from the real axis.}

It would be interesting to derive the behaviour \eqref{eq:opescaling} and the sum rule \eqref{eq:SlowSum} directly from quantum field theory in $\textrm{AdS}_2$. Note that the sum rule is trivially satisfied by the free fermion since then $\alpha=0$, and $1/[\Gamma\left(\frac{1-\delta}{2}\right)\Gamma\left(\frac{3-\delta}{2}\right)]$ vanishes for $\delta$ positive odd integer.

\section{An analytic bound in 2D}\label{sec:2d}
\subsection{The new basis in 2D}
We will now discuss a generalization of the new class of bootstrap functionals to two dimensions and how it can be used to produce an analytic constraint on the low-lying spectrum. The conformal blocks with four identical external scalar primaries $\phi(x)$ read
\be
G_{h,\bar{h}}^{\textrm{2D}}(z,\bar{z}) = G_{h}(z)G_{\bar{h}}(\bar{z}) + (h\leftrightarrow\bar{h})\,,
\ee
where $G_{h}(z)$ is the 1D conformal block \eqref{eq:block1D} and
\be
h = \frac{\Delta+l}{2}\,,\qquad \bar{h} = \frac{\Delta-l}{2}\,,
\ee
where $\Delta$, $l$ are the dimension and spin of the propagating primary. The bootstrap equation now reads
\be
\sum\limits_{\mathcal{O}\in\phi\times\phi}(c_{\phi\phi\mathcal{O}})^2F_{h,\bar{h}}(z,\bar{z}) =0\,,
\label{eq:bs2D}
\ee
where
\be
F_{h,\bar{h}}(z,\bar{z}) =\left[ g_{h}(z)g_{\bar{h}}(\bar{z})-
g_{h}(1-z)g_{\bar{h}}(1-\bar{z})\right] + (h\leftrightarrow\bar{h})\,,
\label{eq:F2D}
\ee
where
\be
g_h(z) = z^{h-\Delta_\phi}{}_2F_1(h,h;2h;z)\,.
\ee
$z,\bar{z}$ should be thought of as independent complex variables. For any value of $\bar{z}$, functions $F_{h,\bar{h}}(z,\bar{z})$ have a pair of branch cuts in $z$ located at $z\in(-\infty,0)$ and $z\in(1,\infty)$, and vice versa with $z$ and $\bar{z}$ interchanged. Let us then define a basis of linear functionals $\alpha_n$ acting on functions $g_{h}(z)$ as in subsection \ref{ssec:newbasis}
\bee
\alpha_n\left[g_h(z)\right] &= s^+(h,n)\\
\alpha_n\left[g_h(1-z)\right] &= s^-(h,n)\,,
\eee
where $s^{\pm}(h,n)$ appear in \eqref{eq:sPlus} and \eqref{eq:sMinus}. It is also convenient to define basis functionals $\beta_n$ acting in the opposite way, i.e. by scalar products of Legendres against the discontinuity on the branch cut $z\in (-\infty,0)$
\bee
\beta_n\left[g_h(z)\right] &= s^-(h,n)\\
\beta_n\left[g_h(1-z)\right] &= s^+(h,n)\,.
\eee
These functionals are not independent from $\alpha_n$ in 1D, but are needed in 2D. The basis for our class of functionals acting on the 2D crossing equation consists of the following tensor products
\bee
(\alpha_m\otimes\bar{\alpha}_n)\left[F_{h,\bar{h}}(z,\bar{z})\right] &= \left[s^+(h,m)s^+(\bar{h},n)-s^-(h,m)s^-(\bar{h},n)\right] + (h\leftrightarrow \bar{h})\\
(\alpha_m\otimes\bar{\beta}_n)\left[F_{h,\bar{h}}(z,\bar{z})\right] &= \left[s^+(h,m)s^-(\bar{h},n)-s^-(h,m)s^+(\bar{h},n)\right] + (h\leftrightarrow \bar{h})\,,
\label{eq:2DBasAct}
\eee
where $\alpha,\beta$ acts on the $z$ variable, while $\bar{\alpha},\bar{\beta}$ acts on the $\bar{z}$ variable. The other combinations are not independent since
\bee
\beta_m\otimes\bar{\beta}_n &= - \alpha_m\otimes\bar{\alpha}_n\\
\beta_m\otimes\bar{\alpha}_n &= - \alpha_m\otimes\bar{\beta}_n
\eee
when acting on $F_{h,\bar{h}}(z,\bar{z})$. The symmetrization under $h\leftrightarrow\bar{h}$ in $F_{h,\bar{h}}(z,\bar{z})$ guarantees that the functionals appearing in \eqref{eq:2DBasAct} satisfy symmetry properties
\bee
\alpha_n\otimes\bar{\alpha}_m &= \alpha_m\otimes\bar{\alpha}_n\\
\alpha_n\otimes\bar{\beta}_m &= - \alpha_m\otimes\bar{\beta}_n\,.
\eee
It is therefore natural to use the following as an independent basis of functionals with $m,n\in\mathbb{N}$
\be
\gamma_{mn} = \alpha_m\otimes\bar{\alpha}_n + \alpha_m\otimes\bar{\beta}_n\,,
\ee
so that the first, second line of \eqref{eq:2DBasAct} are respectively the symmetric and antisymmetric part of the matrix $\gamma_{mn}$. The action of $\gamma_{mn}$ becomes
\be
\gamma_{mn}\left[F_{h,\bar{h}}(z,\bar{z})\right] = s(h,m)\tilde{s}(\bar{h},n)+ (h\leftrightarrow \bar{h})\,,
\ee
where
\bee
s(h,m) &= s^+(h,m)-s^-(h,m)\\
\tilde{s}(h,m) &= s^+(h,m)+s^-(h,m)\,.
\eee
In other words, $s(h,m)$ is precisely the function \eqref{eq:s1D} giving the action of $\alpha_n$ on the vectors $F_h(z)$ entering the crossing equation in 1D. On the other hand, $\tilde{s}(h,m)$ is the action of $\alpha_n$ on the vectors $\tilde{F}_{\Delta}(z)$ for the 1D crossing equation with the wrong sign
\bee
&\sum\limits_{\mathcal{O}\in\psi\times\psi}(c_{\psi\psi\mathcal{O}})^2\tilde{F}_{\Delta_{\mathcal{O}}}(z)=0\\
&\tilde{F}_{\Delta}(z) = z^{-2\Delta_\psi}G_{\Delta}(z) + (1-z)^{-2\Delta_\psi}G_{\Delta}(1-z)\,.
\label{eq:1DBootPlus}
\eee
It is easy to see this equation has no nontrivial solution, as witnessed by any functional in the form of a positive linear combination of even derivatives of $\left[z(1-z)\right]^{2\Delta_\psi}\tilde{F}_{\Delta}(z)$ evaluated at $z=1/2$.

\subsection{Analytic bounds from factorized functionals}
Having defined a natural basis for conformal bootstrap functionals in 2D, the remaining task is to find coefficients $a_{mn}\in\mathbb{R}$ so that
\be
\omega = \sum_{m,n\in\mathbb{N}}a_{mn}\gamma_{mn}
\ee
is an extremal functional. Since the action of $\gamma_{mn}$ essentially factorizes into a holomorphic and antiholomorphic part, it is natural to consider a restricted class of functionals where $a_{mn}$ factorizes into a pair of sequences
\be
a_{mn} = a_{m}\tilde{a}_{n}\,.
\ee
The action of $\omega$ then becomes
\be
\omega(F_{h,\bar{h}}) = u(h)\tilde{u}(\bar{h}) + u(\bar{h})\tilde{u}(h)\,,
\label{eq:omAct2D}
\ee
where
\bee
u(h) &= \sum\limits_{m\in\mathbb{N}} a_m s(h,m)\\
\tilde{u}(h) &= \sum\limits_{n\in\mathbb{N}} \tilde{a}_n \tilde{s}(h,n)\,.
\eee
We will take $a_n$ to be the coefficients in the 1D extremal functional for the gap $2\Delta_\psi + 1=\Delta_\phi + 1$, assuming it exists for any $\Delta_\phi$. Hence
\bee
&u(0) = 0\\
&u(h)\textrm{ has a first-order zero and a positive slope at }h=\Delta_\phi+1\\
&u(h)\geq 0\textrm{ for }h\geq\Delta_\phi + 1\\
&u(h)\textrm{ has second-order zeros at }h=\Delta_\phi+2j + 1\,,j\in\mathbb{N}\,.
\eee
Since the 1D crossing with the wrong sign has no nontrivial solutions, it is easy to find $\tilde{a}_n$ such that
\be
\tilde{u}(h) > 0
\ee
for all $h\geq 0$. We can take for example $\tilde{a}_n$ corresponding to the functional in the form of the second derivative of $\left[z(1-z)\right]^{2\Delta_\psi}\tilde{F}_{\Delta}(z)$ at $z=1/2$. Consider now \eqref{eq:omAct2D} as a function of $\Delta$ for fixed $l$, denoting $\omega(\Delta,l) = \omega(F_{h,\bar{h}})$. We find the following properties
\bee
&\omega(0,0) = 0\\
&\omega(\Delta,l)\textrm{ has a first-order zero and a positive slope at }\Delta=2\Delta_\phi+2+l\,,\,l\in2\mathbb{N}\\
&\omega(\Delta,l)\geq 0\textrm{ for }\Delta\geq 2\Delta_\phi+2+l\,,\,l\in2\mathbb{N}
\eee
It follows that unless all primary operators of a unitary solution to \eqref{eq:bs2D} coincide with the zeros of $\omega(\Delta,l)$, the solution must contain at least one primary in the negative region of $\omega(\Delta,l)$. In other words, there must be a primary distinct from identity with twist
\be
\tau_{\textrm{gap}}\leq 2\Delta_\phi + 2\,.
\label{eq:TwistBound}
\ee
An analogous result holds in $d>2$, where there must exist an operator with $\tau$ arbitrarily close to $2\Delta_\phi$ \cite{LargeSpin1,LargeSpin2}. This upper bound on the minimal twist does not rely on Virasoro symmetry, and therefore holds for arbitrary 2D conformal defects. In fact, Virasoro symmetry implies the existence of operators with $\tau = 0$ and $l\geq 2$, so the bound is automatically satisfied. However, the functional $\omega$ carries useful information in this case too. Note that when $0\leq \Delta_\phi\leq 1$
\be
\omega(l,l) = u(l)\tilde{u}(0) \geq 0\quad\textrm{for }l\in 2\mathbb{N}\,,
\ee
so the Virasoro descendants of identity do not help and the bound \eqref{eq:TwistBound} must be satisfied by an operator of non-zero twist. Assuming the operator of minimal non-zero twist is a scalar (such as in all theories with Virasoro symmetry where all Virasoro primaries are scalar), numerical bootstrap shows our analytic bound is strictly above the optimal upper bound on the scalar gap for $0<\Delta_\phi<1$, but it becomes optimal at $\Delta_\phi = 1$, where the extremal solution corresponds to the correlator $\langle \epsilon\epsilon\epsilon\epsilon \rangle$ in the 2D Ising model, the twist-four primary being $\mathcal{L}_{-2}\bar{\mathcal{L}}_{-2}\mathds{1}$.

\section{Future directions}\label{sec:future}
We expect that the class of functionals introduced in this work will be useful for extracting analytic predictions from the conformal bootstrap equations in a wider variety of contexts. Work is currently in progress to generalize the results to more spacetime dimensions, where the bootstrap bounds exhibit interesting features at locations corresponding to interacting CFTs.

An especially promising property of our functionals is their well-controlled behaviour when the external scaling dimensions are large. This is in sharp contrast with the derivative functionals normally used for the numerical bootstrap, whose constraining power deteriorates with increasing external dimensions \cite{SMatrix1}. For this reason, the functionals from this paper are useful for extracting the consequences of boundary crossing symmetry on $\textrm{AdS}$ physics, as demonstrated in section \ref{sec:ads}. In this context, it would also be interesting to test the sum rule \eqref{eq:SlowSum}, for example by constructing exact solutions to crossing corresponding to scattering in integrable theories in large $\textrm{AdS}_2$.

It would also be very interesting to identify the extremal functionals for bounds on OPE coefficients analytically in our basis. In 1D with large external scaling dimension, the numerical upper bound on the OPE coefficients of bound states coming from CFT crossing was observed to coincide with the corresponding analytical bound coming from S-matrix bootstrap in flat space \cite{SMatrix1,SMatrix2}. It is conceivable that similar methods to those presented in our paper can be used to prove this upper bound analytically on the CFT side. The main challenge seems to be able to place the double zeros of the extremal functionals to more general locations than the equally spaced points occuring in the present work.

Crossing symmetry of mixed correlators dramatically improves the bootstrap bounds \cite{3dIsingMixed, ONMixed}. Similar improvements are observed to occur in 1D \cite{1DMixed}. Our basis for functionals is expected to generalize to this context too, hopefully paving the way towards an analytic understanding of bootstrap islands in more spacetime dimensions. Related functionals might also be useful for modular bootstrap, where suggestions concerning the analytic nature of extremal functionals recently appeared in \cite{ModularBootstrap}.

It could be fruitful to explore the utility of our basis for standard numerical bootstrap. Truncating the space of integral kernels to the span of Legendre polynomials of a bounded order, it is impossible to impose positivity for arbitrarily large scaling dimension since the functionals eventually become oscillating. However, one could try imposing positivity only up to the maximum order of a Legendre used.

Finally, one should look for an interpretation of what the presented functionals are trying to do physically. The partial restoration of the $z\leftrightarrow 1-z$ symmetry, described in the last paragraph of section \ref{sec:extremalfunctionals} is a hint that our basis, which breaks this symmetry explicitly, might not be the optimal choice.

\section*{Acknowledgements}
I would like to thank Miguel Paulos for many useful discussions and collaboration during an initial stage of this work. I thank Nikolay Bobev, Shira Chapman, Juan Maldacena, Marco Meineri, Leonardo Rastelli, David Simmons-Duffin, Xi Yin and especially Davide Gaiotto and Pedro Vieira for useful discussions, suggestions and comments on the draft.

I thank the organizers of the workshop on Conformal Field Theories and Renormalization Group Flows held in the Galileo Galilei Institute in Florence in the spring 2016 for creating a stimulating environment. 

This research was supported in part by Perimeter Institute for Theoretical Physics. Research at Perimeter Institute is supported by the Government of Canada through the Department of Innovation, Science and Economic Development Canada and by the Province of Ontario through the Ministry of Research, Innovation and Science.
\newpage

\begin{appendices}
\section{Closed formulas for the integral kernel}\label{app:closed}
The goal of this appendix is to explain how one can obtain closed formulas for the integral kernel $\tilde{h}(x)$ corresponding to $\Delta_\psi\in\mathbb{N}-1/2$, specified by the formulas \eqref{eq:hXInteger2} and \eqref{eq:hHalfInteger}. Define
\be
\left\langle f,g\right\rangle = \int\limits_0^1\!\d x f(x)g(x)
\ee
the usual scalar product of real functions on the unit interval. The basis functions are orthogonal
\be
\left\langle p_{m},p_{n}\right\rangle = \frac{\delta_{mn}}{2m-1}\quad\textrm{for }m,n\in\mathbb{N}\,,
\ee
so that $\tilde{h}(x)$ is the unique function satisfying
\be
\tilde{a}_n = (2n-1)\langle\tilde{h},p_n\rangle\,.
\ee
Our strategy for finding $\tilde{h}(x)$ will be to write $\tilde{a}_n/(2n-1)$ as a linear combination of overlaps between $p_n$ and some relatively simple functions. It is useful to define the following functions
\bee
q_{\Delta}(x) &= Q_{\Delta-1}(2x-1)\\
r_{\Delta}(x) &= \frac{\Gamma(\Delta)^2}{2\Gamma(2\Delta)}\left[x^{\Delta-1}{}_2F_1(\Delta,\Delta;2\Delta;x)+ (1-x)^{\Delta-1}{}_2F_1(\Delta,\Delta;2\Delta;1-x)\right]\,,
\eee
where $Q_n(y)$ is the Legendre function of the second kind. When $\Delta$ is even, we have $q_{\Delta}(1-x) = q_{\Delta}(x)$, and therefore $\langle q_\Delta,p_n\rangle$ is nonzero only for $n$ odd. In that case, we find the following overlaps
\be
\langle q_\Delta,p_n\rangle = \frac{1}{\lambda-\Delta(\Delta-1)}\,,
\label{eq:qpoverlap}
\ee
where $\lambda = n(n-1)$. Clearly, the overlaps $\langle r_\Delta,p_n\rangle$ are nonvanishing again only for $n$ odd. For $\Delta=1,2$ we find
\bee
\langle r_1,p_n\rangle &= -\frac{1}{2}\Psi'\left(\frac{n}{2}\right)\\
\langle r_2,p_n\rangle &= \left[n(n-1)+\frac{1}{2}\right]\Psi'\left(\frac{n}{2}\right) + 2\,,
\eee
with $\Psi(z)=\psi(z+1/2)-\psi(z)$, $\psi(z)=\Gamma'(z)/\Gamma(z)$. It is now possible to see that
\be
\frac{\tilde{a}_n}{2n-1} = - \langle r_2,p_n\rangle +(2\Delta_\psi - 1)\langle r_1,p_n\rangle+H_{\Delta_\psi}(\lambda)\,,
\ee
where $H_{\Delta_\psi}(\lambda)$ is a rational function of $\lambda$ with simple and double poles at $\lambda = \Delta(\Delta-1)$ where $\Delta\in\{2,4,\ldots,2\Delta_\psi+1\}$. $H_{\Delta_\psi}(\lambda)\rightarrow0$ as $\lambda\rightarrow\infty$ and therefore $H_{\Delta_\psi}(\lambda)$ can be written as a linear combination of
\be
\frac{1}{[\lambda - \Delta(\Delta-1)]^a}
\ee
with $a=1,2$ and $\Delta\in\{2,4,\ldots,2\Delta_\psi+1\}$. Any summand with $a=1$ can be written as the overlap $\langle q_\Delta, p_n\rangle$ thanks to \eqref{eq:qpoverlap}. It remains to find functions $\tilde{q}_{\Delta}(x)$ such that
\be
\langle \tilde{q}_{\Delta},p_n\rangle = \frac{1}{[\lambda- \Delta(\Delta-1)]^2}\,.
\ee
We could not find a closed formula for $\tilde{q}_\Delta(x)$ but worked out a few low-lying examples. It was useful to notice that $\tilde{q}_\Delta(x)$ is the solution of the Legendre equation with resonant forcing
\be
\left[x(1-x)\tilde{q}'_{\Delta}(x)\right]'+\Delta(\Delta-1)\tilde{q}_\Delta(x) = -q_\Delta(x)
\ee
and boundary conditions
\be
\tilde{q}_\Delta(0) = \tilde{q}_\Delta(1) = \frac{\psi'\left(\frac{1-\Delta}{2}\right)-\psi'\left(\frac{\Delta}{2}\right)}{4(2\Delta-1)}\,.
\ee
The bottom line is that $\tilde{h}(x)$ can be written as a linear combination of $r_1(x)$, $r_2(x)$, $q_\Delta(x)$ and $\tilde{q}_\Delta(x)$ with $\Delta\in\{2,4,\ldots,2\Delta_\psi+1\}$. In this way, one can obtain the following explicit formulas for small $\Delta_\psi$. The kernel for $\Delta_\psi = 1/2$ takes the form
\be
\tilde{h}(x) = \left[\frac{1-y}{2 y}-\frac{\left(2 x^2+x+2\right) (x-1)}{2 x^2}\log (1-x) \right]+ (x\leftrightarrow 1-x)\,,
\ee
where we use the shorthand notation $y=x(1-x)$. The kernel for $\Delta_\psi = 3/2$ takes the form
\be
\tilde{h}(x) = \left[\frac{12 y^2+11 y+12}{24 y}-\frac{\left(2 x^2+3 x+2\right) (x-1)^3}{2 x^2}\log (1-x)\right]+ (x\leftrightarrow 1-x)\,.
\ee
Beginning from $\Delta_\psi=5/2$, we get contributions from $\tilde{q}_\Delta(x)$ which contain the dilogarithm. The kernel for $\Delta_\psi = 5/2$ takes the form
\bee
\tilde{h}(x) =&
\left[\frac{-120 y^3+154 y^2+641 y+120}{240 y}
+\frac{3}{5}(2 x-1) (y+2)\mathrm{Li}_2(x)+\right.\\
&+\left.
\frac{(x-1)(y-1)\left(10 x^4-5 x^3-22 x^2-5 x+10\right)}{10 x^2}\log (1-x)\right]
+ (x\leftrightarrow 1-x)
\,.
\eee
Note that the full kernel is given by $h(x) = \tilde{h}(x) + c(x)$ with $c(x)$ given by \eqref{eq:cpolys}. It turns out that the kernel has simple transformation properties under $z\leftrightarrow 1-z$. We leave further exploration of these for a future study.

\end{appendices}
\newpage
\bibliographystyle{utphys}
\bibliography{Bootstrap}
\end{document}